\documentclass[11pt,a4paper]{article}

\usepackage[margin=1in]{geometry}
\usepackage{graphicx}
\usepackage{amsmath,amssymb,amsthm}
\usepackage{booktabs}
\usepackage{natbib}
\usepackage{hyperref}
\usepackage{url}
\usepackage{xcolor}
\usepackage[titletoc]{appendix}
\usepackage{enumitem}
\usepackage{pdfpages}

\graphicspath{{Fig/}}

\hypersetup{
  colorlinks=true,
  linkcolor=blue!50!black,
  citecolor=blue!50!black,
  urlcolor=blue!50!black
}

\theoremstyle{plain}
\newtheorem{theorem}{Theorem}
\newtheorem{proposition}[theorem]{Proposition}
\theoremstyle{remark}
\newtheorem{remark}{Remark}
\theoremstyle{definition}

\newcommand{\botrule}{\bottomrule}
\newenvironment{tablenotes}{%
  \par\smallskip\begin{minipage}{\linewidth}\footnotesize
  \renewcommand{\item}[1][]{\par\noindent}%
}{\end{minipage}\par\smallskip}
\newenvironment{unlist}{\begin{description}[style=nextline,leftmargin=1.5em]}{\end{description}}

\title{Modelling Athletic Ageing Relative to an Estimated Performance Envelope}
\author{%
Dae-Jin Lee\\
School of Science and Technology, IE University\\
Cardenal Z\'{u}\~{n}iga S/N, 40003 Segovia, Spain\\
\texttt{Dae-Jin.Lee@ie.edu}\\
ORCID: \href{https://orcid.org/0000-0002-8995-8535}{0000-0002-8995-8535}
}
\date{}

\begin{document}
\maketitle

\begin{abstract}
Athletic careers yield sparse, irregular longitudinal series: few seasons per athlete, incomplete paths, and selection into continued play. Scientific interest often centres on proximity to peak attainable performance---a ceiling---rather than on the mean trajectory, and on how that proximity co-varies with the rate of decline. Standard tools address only pieces of this problem. Linear mixed models describe average ageing; functional principal components describe dominant modes of variation; shape-invariant models register curves about a mean template; growth charts estimate population centiles but stop short of individual latent trajectories. We develop RACE (\emph{Relative Aging Curves via Envelopes}), a two-stage framework that estimates a population performance envelope as an age-conditional high centile and then models each athlete as a low-dimensional geometric transformation of that envelope. STAR (\emph{Shape Translation And Rotation}) is the Stage~2 mixed model, with parameters for level, timing, and tempo. Envelope geometry determines which of these parameters are identifiable when careers are short: near-linear envelopes identify level and tempo only, whereas curved envelopes identify all three. Embedding STAR in a nonlinear mixed-effects hierarchy makes the level--tempo association a parameter of the random-effect covariance rather than a post-hoc correlation of separate fits. Simulations ask whether that association is recoverable under sparsity, how geometry governs identifiability, and how sensitive results are to envelope misspecification. Applied to Major League Baseball Statcast sprint speed and bolt rate, the analysis demonstrates how the proposed embedding separates athletic level and ageing tempo in Functional Ageing Space, and shows why timing is identifiable for bolt rate but not for sprint speed.
\end{abstract}


\vspace{0.8em}
\noindent\fbox{\begin{minipage}{0.97\textwidth}
\textbf{Key Messages}
\begin{itemize}[leftmargin=1.2em]
\item Sparse athletic careers should be read relative to a performance envelope (ceiling), not only the mean trajectory.
\item Envelope geometry governs which individual ageing parameters---level, timing, tempo---are identifiable.
\item The statistical object of interest is each athlete's coordinates in Functional Ageing Space (FAS); the relative longevity index is only a projection of tempo onto a familiar scale.
\end{itemize}
\end{minipage}}
\vspace{1.2em}

\section{Introduction}\label{sec:intro}

How should one characterize individual athletic ageing when careers are short, observations are irregular, and the scientific target is proximity to peak attainable performance rather than the mean? Professional running metrics illustrate the difficulty. Peak sprint speed and related indicators reflect near-maximal capacity \citep{hawley2014,tanakaseals2008,allenhopkins2015}, yet each athlete contributes only a handful of seasons; selection and role changes distort mean curves at older ages \citep{bradbury2009,fair2008}; and applied questions ask how close an athlete sits to the age-specific ceiling and whether higher early level predicts faster subsequent decline.

Existing statistical tools answer related but different questions. Linear mixed models estimate average trajectories and random intercepts or slopes; they do not encode ageing relative to a shared nonlinear performance limit. Functional principal component analysis recovers dominant modes of variation in sparse longitudinal data \citep{yaomullerwang2005,ramsaysilverman2005}, but the modes need not correspond to interpretable level, timing, and tempo relative to a fixed reference. Shape-invariant nonlinear mixed-effects models register individual curves about a \emph{mean} template \citep{lindstrombates1990,kneipgasser1992,kneipramsay2008}; they do not take an upper-tail envelope as the reference, nor do they make explicit how template geometry governs which individual parameters remain identifiable under sparsity. Growth-chart methodology and generalized additive models for location, scale and shape (GAMLSS) centiles \citep{colegreen1992,who2006,rigbystasinopoulos2005} estimate population reference curves $F^{-1}_{Y\mid x}(\tau)$ but stop short of hierarchical individual trajectories and of associations inside a random-effect covariance. Quantile or expectile regressions and stochastic frontier models \citep{koenker2005,neweypowell1987,aigner1977} offer alternative ceilings, yet they do not by themselves deliver a sparse-data Stage~2 geometry for level, timing, and tempo.

The applied problem therefore forces a specific methodological structure. One needs (i)~an interpretable population performance envelope---a functional reference that answers ``how good is attainable at age $x$?''---and (ii)~a low-dimensional model for how individuals deviate from that envelope when only a few observations are available. We call the resulting two-stage approach RACE (\emph{Relative Aging Curves via Envelopes}). Stage~1 estimates the envelope as an age-conditional high centile. Stage~2 models individuals relative to that envelope; STAR (\emph{Shape Translation And Rotation}) is the geometric mixed model used here, with level ($\alpha$), timing ($\gamma$), and tempo ($\delta$). The envelope is not a preprocessing convenience: it is the statistical reference object that defines the individual parameters and determines which of them can be learned from short careers.

Rather than ranking athletes by performance, we embed each longitudinal trajectory into a two-dimensional Functional Ageing Space (FAS) defined by level ($\alpha$) and tempo ($\delta$), with timing ($\gamma$) when the envelope is curved. The statistical object of interest is no longer the trajectory itself, but its coordinates in that ageing space. The hierarchy of the method is deliberate:
\begin{itemize}
\item RACE: two-stage framework relative to an envelope;
\item Stage~1: the population performance envelope;
\item Stage~2: STAR geometry and hierarchical covariance $\Sigma$;
\item STAR parameters $(\alpha,\gamma,\delta)$: identifiable coordinates of individual ageing;
\item Functional Ageing Space (FAS): the identifiable STAR coordinates relative to the fitted envelope;
\item relative longevity index (RLI): a derived projection of tempo onto a familiar scale---not the methodology.
\end{itemize}

Two statistical consequences follow directly from athletic sparsity. First, \emph{geometry governs identifiability}: under a near-linear envelope, timing and level are confounded, so the Stage~2 model must reduce to level and tempo; under clear curvature, all three parameters are recoverable. Second, scientifically central associations---notably the correlation between level and tempo---should be estimated as parameters of a hierarchical covariance $\Sigma$, not as sample correlations of separate player-by-player fits, which are biased under short series.

We demonstrate the framework on Major League Baseball Statcast sprint speed (continuous; near-linear envelope) and bolt rate (unit-interval; curved envelope). The application is chosen because the two metrics stress-test the geometric message under different likelihoods. Beyond method recovery, the analysis maps athletes in FAS, documents a strong negative level--tempo association inside $\Sigma$, and shows why timing is not identifiable for sprint speed but is for bolt rate.

Section~\ref{sec:framework} states the RACE framework, STAR, FAS, and identifiability. Section~\ref{sec:related} situates RACE relative to existing methodology. Section~\ref{sec:estimation} describes hierarchical estimation. Section~\ref{sec:simulation} reports simulations on correlation recovery, geometry and identifiability, and envelope misspecification. Section~\ref{sec:application} presents the Statcast analysis. Section~\ref{sec:discussion} discusses limitations and extensions.

\section{The RACE framework}\label{sec:framework}

\subsection{Envelope as the statistical reference}\label{sec:ceiling-def}

In athletic ageing the scientifically natural reference is often an upper functional limit: at each age, how fast or how often do the best performers run? We take the population envelope to be an age-conditional high centile,
\begin{equation}\label{eq:envelope}
f(x)=F^{-1}_{Y\mid x}(\tau),\qquad \tau\in(0,1),
\end{equation}
with $\tau=0.95$ in the applications---written C95 for the age-conditional $95$th centile---so that roughly $5\%$ of athlete-seasons exceed the ceiling at each age. Modelling deviations from $f$ answers a different question from modelling deviations from $E(Y\mid x)$: the former concerns capacity relative to what is attainable; the latter concerns average behaviour, which selection into continued play can distort \citep{bradbury2009}.

We estimate~\eqref{eq:envelope} by GAMLSS \citep{rigbystasinopoulos2005,rigby2019}, fitting
$Y\mid x\sim\mathcal{D}\bigl(\mu(x),\sigma(x),\ldots\bigr)$
with smooth age effects on the distributional parameters and extracting the fitted quantile function. The target is a growth-chart centile in the LMS/WHO sense \citep{colegreen1992,who2006}: a population ceiling against which sparse careers are later assessed. GAMLSS is the natural distributional-regression engine for that object \citep{rigbystasinopoulos2005,kneib2013}.

Four features matter for athletic envelopes and are not delivered together by a single semiparametric quantile fit: age-varying scale (and shape) that borrows strength in a thin upper tail; family choice matched to support (positive continuous sprint; unit-interval bolt with opportunity weights); a monotone-decreasing mean constraint \emph{inside} the observation model ($\mathrm{pbm}$, $\mathrm{mono}=\texttt{"down"}$); and a coherent multi-centile fan from one fit (Figure~\ref{fig:envelopes}). Quantile regression targets the same $F^{-1}(\tau\mid x)$ more directly and is a sensitivity check; expectiles are a different functional (Supplementary Material, Section~E). We keep Stage~1 and Stage~2 separate for the same reason growth charts are separate from individual assessment: Stage~1 estimates a population ceiling, whereas folding STAR random effects into the centile fit would change that estimand. Section~E also reports P-spline edf and a player-cluster bootstrap that re-estimates $\hat f$ into Stage~2.

\subsection{STAR: geometric Stage~2 for sparse careers}\label{sec:star}

Let $\hat f$ denote the Stage~1 envelope, treated as fixed for estimation of individual parameters. For athlete $i$ at age $x_{ij}$, STAR specifies
\begin{equation}\label{eq:star}
y_i(x)=\alpha_i+\hat f\!\left(\frac{x-\gamma_i}{\exp(\delta_i)}\right),
\end{equation}
with observation models matched to the metric: Gaussian errors for continuous sprint speed, and a Simplex law for bolt rates in $(0,1)$ \citep{barndorffjorgensen1991},
\begin{align}
\text{Sprint:}~ &
y_{ij}=\alpha_i+\hat f\!\left(\frac{x_{ij}}{\exp(\delta_i)}\right)+\varepsilon_{ij},\quad
\varepsilon_{ij}\sim\mathrm{Normal}(0,\sigma_\varepsilon^2),
\label{eq:sprint}
\\
\text{Bolt:}~ &
Y_{ij}\sim\mathrm{SIMPLEX}(\mu_{ij},\sigma),\quad
\mathrm{logit}(\mu_{ij})=\alpha_i+\hat f\!\left(\frac{x_{ij}-\gamma_i}{\exp(\delta_i)}\right),
\label{eq:bolt}
\end{align}
the latter weighted by competitive opportunities as in Stage~1. Equation~\eqref{eq:sprint} omits timing because the sprint envelope is approximately linear (Section~\ref{sec:ident}). The maps have direct athletic readings (Table~\ref{tab:star-params}).
\begin{table}[t]
\centering
\caption{STAR individual parameters and athletic interpretation.\label{tab:star-params}}%
\begin{tabular*}{\linewidth}{@{\extracolsep\fill}llp{0.52\linewidth}@{\extracolsep\fill}}
\toprule
Symbol & Name & Athletic interpretation \\
\midrule
$\alpha_i$ & Level & Vertical shift relative to the ceiling ($\alpha_i<0$: below peak attainable) \\
$\gamma_i$ & Timing & Horizontal age shift (earlier/later ageing schedule) \\
$\delta_i$ & Tempo & Log time-scale: $\delta_i>0$ slows decline; $\delta_i<0$ accelerates it \\
\botrule
\end{tabular*}
\begin{tablenotes}
\item Source: Structural mean~\eqref{eq:star}; see also Figure~\ref{fig:star}.
\end{tablenotes}
\end{table}
When $\hat f$ is locally linear with slope $b<0$, the effective annual decline is $b_i^*=b e^{-\delta_i}$. The envelope $\hat f$ enters Stage~2 as a fine age grid with linear interpolation; careers that push transformed ages outside the observed range use edge-slope extrapolation on a padded grid (Supplementary Material).

\begin{remark}[Interpretable reparameterization]\label{rem:reparam}
Equation~\eqref{eq:star} is the structural mean used for estimation; display need not use $(\alpha,\gamma,\delta)$ alone. Under a locally linear envelope one may report the equivalent deficit--slope form
\[
\mu_i(x)=L_i+S_i(x-x_0),\qquad
L_i=\hat f(x_0)+\alpha_i,\qquad
S_i=b\,e^{-\delta_i}=b_i^*,
\]
so that $L_i$ is level at a reference age $x_0$ and $S_i$ is absolute annual decline (e.g.\ ft/sec per year). Equivalently, $\lambda_i=e^{\delta_i}$ reads as calendar years per unit of envelope age ($\lambda_i>1$: slower ageing). For a curved envelope with peak age $x^\star$ of $\hat f$, the individual peak $x_{\mathrm{peak},i}=\gamma_i+x^\star e^{\delta_i}$ is often clearer than raw $(\gamma_i,\delta_i)$. These are one-to-one transforms of STAR parameters, not alternative Stage~2 models; we keep~\eqref{eq:star} for fitting and geometry. A tempo display relative to $\hat\mu_\delta$ appears later as a derived projection (Section~\ref{sec:rli-display}), not as part of the Stage~2 estimand.
\end{remark}

STAR is a shape-invariant nonlinear mixed-effects model in the sense of \citet{lindstrombates1990}, with structural mean given by the envelope composed with the maps in~\eqref{eq:star}, population means as fixed effects, and individual deviations as random effects. Relative to classical shape-invariant NLME about a mean curve, the template is an upper-tail performance envelope and an explicit level shift is retained---both required by the athletic question.

The STAR parameters define the \emph{Functional Ageing Space} (FAS): each athlete is a point $(\alpha_i,\delta_i)$ when timing is fixed, or $(\alpha_i,\gamma_i,\delta_i)$ when the envelope is curved. FAS is simply the identifiable random-effect space of a specific STAR model relative to a specific envelope---analogous in role to a SITAR or latent growth space---not a claim of a universal ageing geometry. Population structure in FAS is read from the hierarchical law of the random effects and from the empirical distribution of BLUPs; displays such as density contours in Section~\ref{sec:fas} visualise that object.

\begin{figure}[tbp]
\centering
\includegraphics[width=0.85\textwidth]{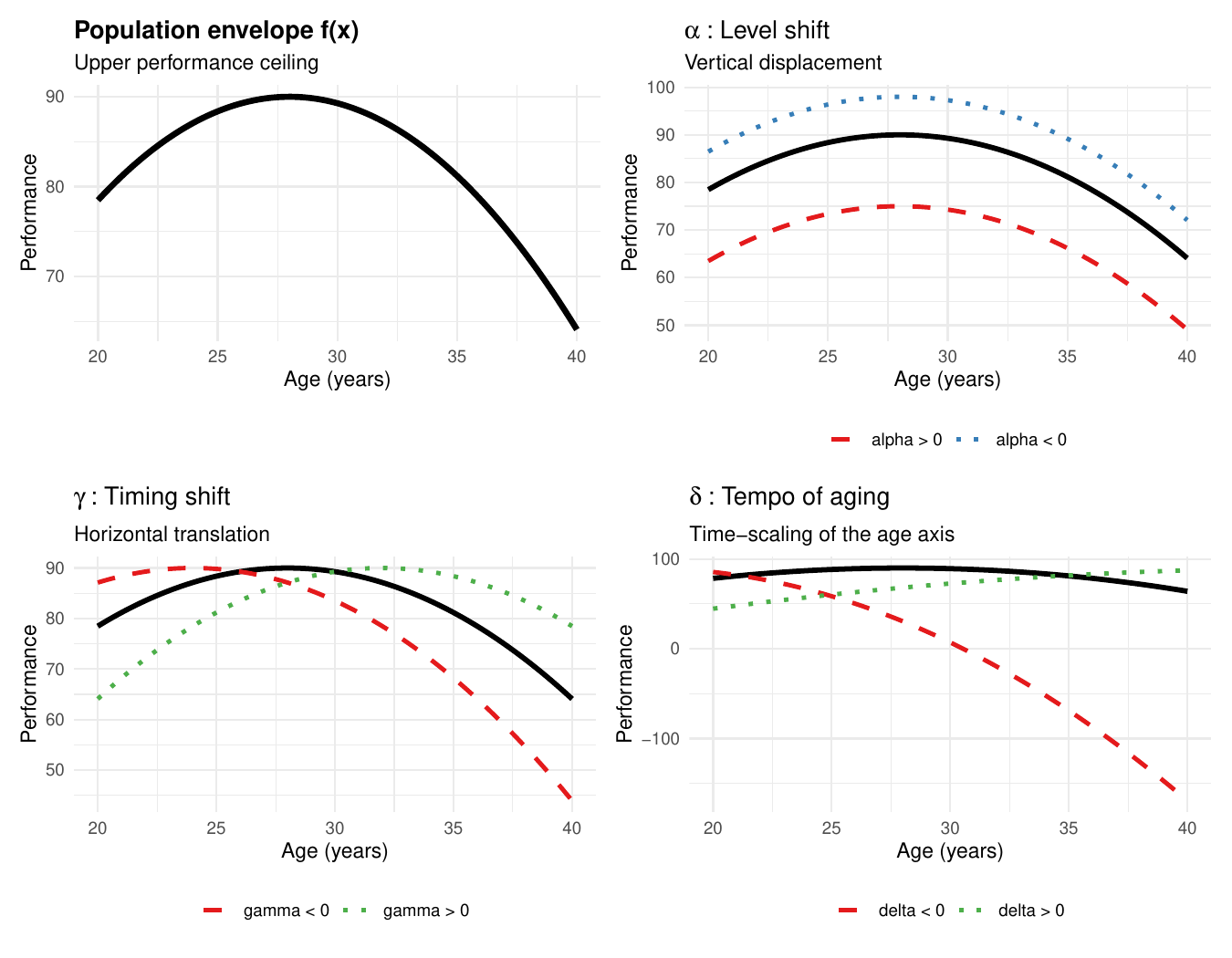}
\caption{STAR transformations of a population envelope: level ($\alpha$), timing ($\gamma$), and tempo ($\delta$). Solid: envelope; dashed/dotted: transformed individual trajectories.\label{fig:star}}
\end{figure}

\subsection{Identifiability and envelope geometry}\label{sec:ident}

Sparse careers cannot support an arbitrarily rich individual parameter vector. The geometry of $f$ decides what is recoverable.

\begin{proposition}[Linear envelopes]\label{prop:linear}
If $f(x)=a+bx$ with $b\neq 0$, then
\[
\alpha+f\!\bigl((x-\gamma)e^{-\delta}\bigr)
=
\bigl(\alpha+a-b\gamma e^{-\delta}\bigr)+b e^{-\delta}\,x.
\]
Only an effective intercept and an effective slope are identified: $\alpha$ and $\gamma$ are confounded. In particular, timing is not separately recoverable from level.
\end{proposition}

\begin{proposition}[Curved envelopes]\label{prop:curved}
Suppose $f$ is twice continuously differentiable on an open interval containing the transformed ages of interest, with $f''$ not identically zero there, and the observation design places at least three distinct ages in that interval. Then, for known $f$ and for true $\delta$ in a neighbourhood of $0$, $(\alpha,\gamma,\delta)$ are locally identifiable from the structural mean~\eqref{eq:star} at that value.
\end{proposition}

\noindent
Proofs and regularity details appear in Supplementary Material, Section~A. Proposition~\ref{prop:linear} is an \emph{exact}-linearity statement; Proposition~\ref{prop:curved} is a \emph{local} injectivity statement near $\delta=0$. Neither alone justifies the applied binary choice ``near-linear $\Rightarrow$ fix $\gamma$; curved $\Rightarrow$ free $\gamma$'' used for sprint versus bolt. Two bridges are needed.

\paragraph{From exact to practical near-linearity.}
Real envelopes are never exactly linear. What matters under sparse noisy careers is whether a timing shift is distinguishable from a level shift at the residual scale. Write $m(x;\theta)=\alpha+f((x-\gamma)e^{-\delta})$ and let $J(x)$ be the Jacobian of $m$ in $(\alpha,\gamma,\delta)$ at $\delta=0$, $\gamma=0$. For an exactly linear $f$, the first two columns of $J$ are linearly dependent at every $x$ (Proposition~\ref{prop:linear}). For a mildly curved $f$, they are formally independent, but the smallest singular value of a stacked design $J(x_1),\ldots,J(x_n)$ can still be tiny relative to the observation noise. A simple operational check on the fitted Stage~1 C95 over ages $20$--$40$ is the residual root-mean-square of a linear fit to $\hat f$, together with the size of a one-year timing perturbation relative to the Stage~2 residual scale. For sprint speed the linear approximation leaves RMS residual $\approx 0.011$~ft/sec against Stage~2 $\hat\sigma\approx 0.42$, and a one-year timing shift has signal-to-noise ratio $\approx 0.3$ on a four-age probe design: timing is theoretically free once $f''\not\equiv 0$, but practically unrecoverable, so we fix $\gamma_i=0$. Bolt rate's C95 on the probability scale is materially less linear (linear $R^2\approx 0.98$ versus $>0.999$ for sprint); Stage~2 on the logit-STAR mean recovers a positive $\hat\omega_\gamma$ (Section~\ref{sec:geometry-insight}). Details and the Jacobian construction are in Supplementary Material, Section~A.

\paragraph{From local to applied claims.}
Proposition~\ref{prop:curved} does not assert global identifiability over the full $(\alpha,\gamma,\delta)$ domain used in Table~\ref{tab:compare}. The sprint/bolt contrast in the application is therefore an empirical consequence of envelope geometry under the fitted Stage~2 models, not a corollary of local injectivity alone. Simulation Question~2 (Section~\ref{sec:simulation}) supplies the missing bridge: under exact linear truth, $\gamma_i$ collapses to a fixed reference; under quadratic and Gompertz truth with matched sparsity, $\gamma_i$ is recovered. We treat the propositions as geometry diagnostics that justify the Stage~2 parameterisation, and the simulation as the finite-sample warrant for reading Table~\ref{tab:compare} as more than a local statement.

\begin{figure}[tbp]
\centering
\includegraphics[width=0.95\textwidth]{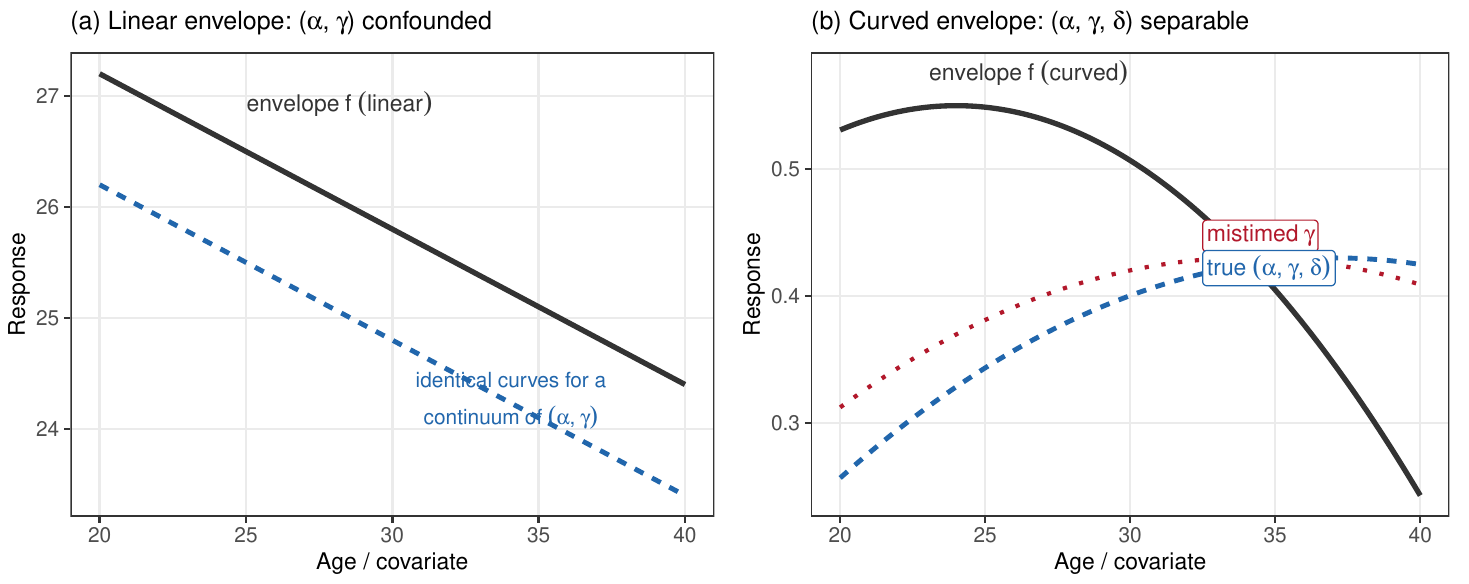}
\caption{Geometry governs identifiability under sparse sampling.
(a)~Linear envelope (solid) and a STAR individual curve (dashed). For linear $f$, a continuum of $(\alpha,\gamma)$ pairs yields the same trajectory: level and timing are confounded, so only an effective intercept and slope are identified.
(b)~Curved envelope (solid) and two STAR trajectories that share the same $(\alpha,\delta)$: the true timing $\gamma$ (dashed) and a counterfactual mistimed $\gamma$ (dash-dotted). The mistimed path is not a defect of STAR; it is a didactic probe showing that, under curvature, a timing error is visible in the mean curve, so $(\alpha,\gamma,\delta)$ separate.
\label{fig:ident}}
\end{figure}

The practical message for athletic data is then summarized in Table~\ref{tab:ident-geom} and Figure~\ref{fig:ident}:
\begin{table}[t]
\centering
\caption{Envelope geometry and identifiable STAR parameters under sparse careers.\label{tab:ident-geom}}%
\begin{tabular*}{\linewidth}{@{\extracolsep\fill}lll@{\extracolsep\fill}}
\toprule
Envelope geometry & Identifiable parameters & Example metric \\
\midrule
Near-linear (practically) & $\alpha_i$, $\delta_i$ ($\gamma_i$ fixed) & Sprint speed \\
Materially curved & $\alpha_i$, $\gamma_i$, $\delta_i$ & Bolt rate \\
\botrule
\end{tabular*}
\begin{tablenotes}
\item Source: Propositions~\ref{prop:linear}--\ref{prop:curved} and the practical near-linearity diagnostic in Section~\ref{sec:ident}; finite-sample support from Simulation Question~2.
\end{tablenotes}
\end{table}

Figure~\ref{fig:ident} is a schematic of Propositions~\ref{prop:linear}--\ref{prop:curved}, not a critique of the STAR formulation~\eqref{eq:star}. Both coloured curves in each panel are valid STAR maps; what changes is whether distinct parameter triples produce distinct trajectories. Panel~(a) shows why timing is not separately recoverable under an exactly linear ceiling, and why a \emph{near}-linear ceiling behaves the same way in practice once noise swamps the weak curvature signal. That is why sprint Stage~2 fixes $\gamma_i=0$ and retains $(\alpha_i,\delta_i)$---a deliberate reduction under Proposition~\ref{prop:linear} and the practical diagnostic above, not a modelling failure. Panel~(b) keeps $(\alpha,\delta)$ fixed and varies only $\gamma$: under material curvature the mistimed path cannot be absorbed by level or tempo, so all three parameters can be separated when enough distinct ages are observed. The sprint--bolt contrast in the application is therefore a geometric fact about the envelope, not merely a difference in $n_i$, and not evidence against STAR.

\begin{remark}[Why this matters for ageing]
Sprint speed declines nearly linearly over ages 20--40 at the residual scale of Stage~2, so asking for a separate ``timing of ageing'' parameter is not an estimation failure---it is the content of Proposition~\ref{prop:linear} made operational. Bolt rate's curved ceiling makes timing meaningful (Proposition~\ref{prop:curved}, with finite-sample support from Simulation Question~2). Envelope shape is therefore part of the scientific model, not a tuning detail.
\end{remark}

\subsection{Hierarchical associations in Functional Ageing Space}\label{sec:hierarchy}

Athlete-level parameters are random effects from a multivariate normal (MVN),
$(\alpha_i,\delta_i)^\top\sim\mathrm{MVN}(\mu,\Sigma)$
(or $(\alpha_i,\gamma_i,\delta_i)^\top$ when timing is present), with random-effect standard deviations $\omega_\alpha,\omega_\delta$ (and $\omega_\gamma$ when timing is present) and level--tempo correlation $\rho_{\alpha\delta}$ as parameters of $\Sigma$; we reserve $\tau$ for the Stage~1 quantile level in~\eqref{eq:envelope} to avoid overloading the symbol. This is the scientifically central association for athletic ageing: do higher-level athletes decline faster? Estimating $\rho_{\alpha\delta}$ inside the mixed model answers that question under sparsity; correlating separate player fits does not (Section~\ref{sec:simulation}).

\section{Relation to existing methodology}\label{sec:related}

RACE is deliberately close to several established strands; the differences are forced by the athletic problem. Classical shape-invariant models and curve registration \citep{lindstrombates1990,kneipgasser1992,lawton1972,kneipramsay2008,ramsayli1998} align individuals to a \emph{mean} template; STAR keeps that geometric spirit but substitutes a performance envelope and retains an explicit level shift so that ``below ceiling'' is a parameter, with template geometry as an explicit design input (Propositions~\ref{prop:linear}--\ref{prop:curved}). LMS/GAMLSS centiles \citep{colegreen1992,rigbystasinopoulos2005} supply Stage~1; RACE adds Stage~2 hierarchical trajectories and $\Sigma$, which growth charts alone do not deliver. Sparse FPCA \citep{yaomullerwang2005,james2000} reduces dimension well but does not, by itself, yield envelope-relative level, timing, and tempo, nor a direct estimand for $\rho_{\alpha\delta}$ under a shared nonlinear ceiling.

Quantile regression is the natural semiparametric competitor for the same centile estimand~\eqref{eq:envelope}; expectiles and stochastic frontiers \citep{koenker2005,neweypowell1987,aigner1977} are related but distinct ceilings. None replaces GAMLSS as the primary Stage~1 engine: they do not jointly deliver age-varying scale, metric-matched families, an in-model monotone constraint, and a coherent multi-centile fan. We therefore keep GAMLSS as the growth-chart reference and treat quantile/expectile curves as plug-in checks (Supplementary Material, Section~E; Discussion). A conventional random-intercept-and-slope model in age remains a natural predictive baseline, yet it cannot encode tempo relative to a curved envelope or estimate ceiling-relative level.

\section{Estimation}\label{sec:estimation}

Primary Stage~2 fits place a single multivariate normal on the random effects, keeping the STAR mean relative to $\hat f$ and the metric-specific observation model. The Stage~1 envelope is treated as a plug-in, as in growth-chart assessment; Supplementary Material, Section~E records P-spline effective degrees of freedom and a player-cluster bootstrap that refits Stage~1 and Stage~2 together for sprint. Population parameters $(\mu,\Sigma,\sigma)$ are estimated by maximizing a Laplace approximation to the marginal likelihood with automatic differentiation in Template Model Builder \citep{kristensen2016}; athlete summaries are BLUPs shrunk toward that single MVN. The computational engine is not the methodological contribution; it is used because the structural mean is a numeric interpolant of $\hat f$ and because bolt rate uses a weighted Simplex likelihood outside stock Gaussian NLME interfaces \citep{pinheirobates2000}. This model is the estimand for the population level--tempo association $\rho_{\alpha\delta}$ inside $\Sigma$ and for coordinates in Functional Ageing Space (Insights~1--2; Sections~\ref{sec:geometry-insight}--\ref{sec:fas}).

\section{Simulation study}\label{sec:simulation}

Athletic sparsity raises three central statistical questions for RACE/STAR: whether hierarchical estimation recovers the level--tempo association under sparsity; how envelope geometry governs identifiability; and how sensitive Stage~2 is to envelope misspecification. We address each with a targeted experiment. Data are generated from known envelopes (linear, quadratic, Gompertz) with random effects from an MVN at prescribed $\rho_{\alpha\delta}$, $n_i\in\{3,5,8\}$ observations per individual, and residual scales matched to the application. Full factorials and summary tables are in the Supplementary Material (Section~F).

\subsection{Question 1: Can hierarchical estimation recover the level--tempo correlation under sparsity?}

Yes---under a linear envelope matching the sprint geometry. The Laplace NLME recovers $\rho_{\alpha\delta}$ across the sparsity grid, whereas post-hoc correlations of separate athlete fits are unreliable under sparsity (wrong sign or near zero when the true association is moderate and positive). For athletic ageing this means that a strong sample correlation of player-by-player fits is not reliable evidence of a performance--longevity trade-off; the hierarchical estimand is. Summary tables are in Supplementary Material, Section~F.

\subsection{Question 2: How does envelope geometry affect identifiability?}

Proposition~\ref{prop:curved} is local near $\delta=0$; the applied sprint/bolt contrast is not. Simulation Question~2 closes that gap empirically. Under an exactly linear envelope, timing $\gamma_i$ is not recoverable and collapses toward a fixed reference, matching Proposition~\ref{prop:linear}. Under quadratic and Gompertz envelopes with the same sparsity grid, $\gamma_i$ becomes recoverable (positive association with truth; lower RMSE than under linear truth), so the finite-sample behaviour tracks the curved case of Proposition~\ref{prop:curved} beyond the local neighbourhood used in the proof. The simulation therefore warrants reading the application contrast (Section~\ref{sec:geometry-insight}) as geometric rather than as an artefact of sample size alone, while leaving a fully global analytic theorem for future work.

\subsection{Question 3: How sensitive is Stage~2 to envelope misspecification?}

When data are generated from a curved envelope but Stage~2 is fit using a linear template (or the reverse), timing estimates deteriorate and $\hat\rho_{\alpha\delta}$ can be attenuated, while level remains comparatively stable. Under quadratic-truth/linear-fit misspecification, $\hat\gamma_i$ collapses toward the fixed reference---as it must, since a linear working template cannot represent curvature---and $\hat\rho_{\alpha\delta}$ is attenuated relative to the correctly-specified fit, while $\hat\mu_\alpha$ is comparatively robust. The reverse case---curved fit to linear-truth data---similarly leaves level estimation close to nominal while timing becomes poorly determined (inflated standard errors rather than systematic bias), because the true data contain no curvature to anchor $\gamma$. Using an envelope whose qualitative geometry matches the truth---linear versus clearly curved---matters more than modest vertical shifts of a correctly shaped template. Quantitative attenuation and timing error under misspecification are summarized in Supplementary Material, Section~F.

\section{Application: sparse athletic capacity trajectories}\label{sec:application}

\subsection{Data and design of the demonstration}\label{sec:data}

We use MLB Statcast data for seasons 2015--2021 \citep{pettigilani2021,statcastSS,statcastBolt}. Two metrics are chosen to stress-test the geometric message under sparse careers (Table~\ref{tab:data}). \emph{Sprint speed} is peak running velocity (ft/sec) in a one-second window---a continuous measure of peak anaerobic capacity whose upper envelope is nearly linear over ages 20--40. \emph{Bolt rate} is the proportion of competitive runs exceeding $30$~ft/sec---a unit-interval outcome with a curved envelope. Stage~1 uses monotone-decreasing P-splines for location and unrestricted P-splines for scale. Family choice is Box--Cox Cole--Green (original) (BCCGo) for sprint and Simplex for bolt, by the generalized Akaike information criterion (GAIC) among fits with a monotone-decreasing C95 and near-nominal coverage (Appendix~\ref{app:families}); for bolt, Appendix Figure~\ref{fig:bolt-family} shows Simplex preferred to beta (BE) and generalized beta type~I (GB1). Stage~2 uses a Gaussian likelihood for sprint (relative to the BCCGo C95) and a Simplex likelihood for bolt (matching the Stage~1 bolt family).

\begin{table}[t]
\centering
\caption{Data summary for sprint speed and bolt rate (MLB Statcast, 2015--2021).\label{tab:data}}%
\begin{tabular*}{\linewidth}{@{\extracolsep\fill}lcc@{\extracolsep\fill}}
\toprule
Characteristic & Sprint speed & Bolt rate \\
\midrule
Envelope players (obs.) & 1{,}021 (3{,}521) & 427 (967) \\
Age range & 20--40 & 20--40 \\
Envelope & GAMLSS C95 (BCCGo) & GAMLSS C95 (Simplex) \\
STAR players ($n_i\geq 3$) & 610 & 141 \\
Median $n_i$ (IQR) & 5 (3--6) & 4 (3--5) \\
Stage~2 likelihood & Gaussian & Simplex \\
STAR parameters & $(\alpha,\delta)$; $\gamma=0$ & $(\alpha,\gamma,\delta)$ \\
\botrule
\end{tabular*}
\begin{tablenotes}
\item Source: MLB Statcast via Baseball Savant \citep{pettigilani2021,statcastSS,statcastBolt}. Stage~1 family choice in Appendix~\ref{app:families}. Median and IQR of seasons per athlete among Stage~2 players with $n_i\geq 3$ (ages 20--40).
\end{tablenotes}
\end{table}

\begin{figure}[tbp]
\centering
\begin{minipage}{0.48\textwidth}
\centering
\includegraphics[width=\linewidth]{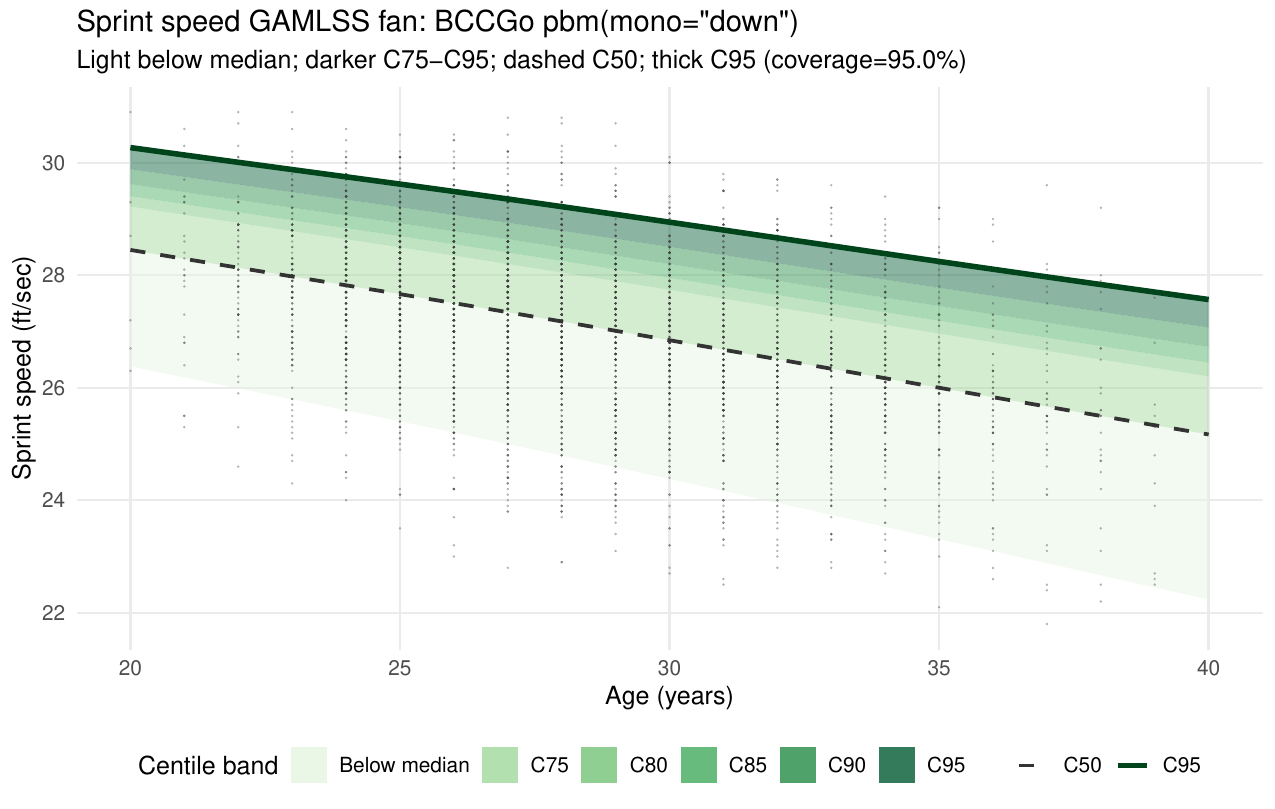}
\end{minipage}\hfill
\begin{minipage}{0.48\textwidth}
\centering
\includegraphics[width=\linewidth]{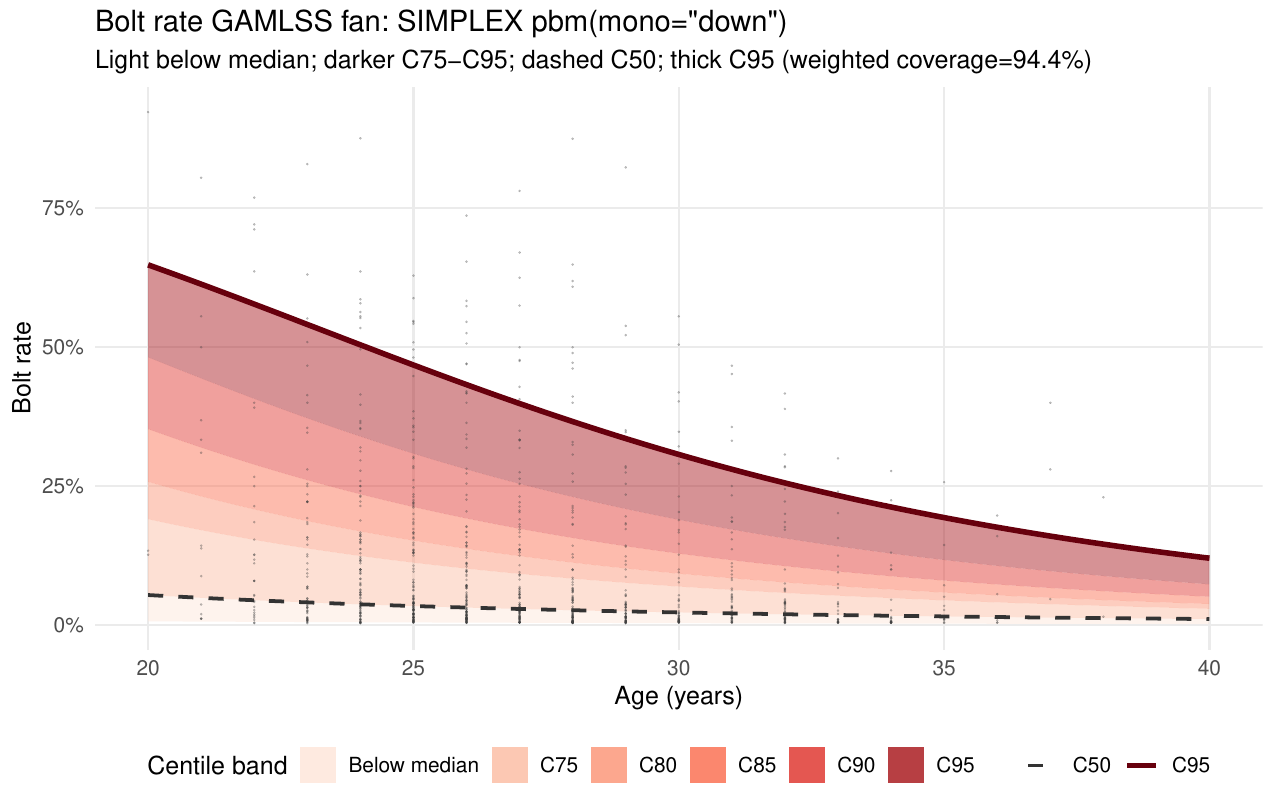}
\end{minipage}
\caption{GAMLSS age-conditional centiles for (a)~sprint speed (BCCGo) and (b)~bolt rate (Simplex). The solid curve is the RACE envelope (C95); lighter curves show lower centiles. Family choice is justified in Appendix~\ref{app:families}.\label{fig:envelopes}}
\end{figure}

\begin{figure}[tbp]
\centering
\includegraphics[width=0.90\textwidth]{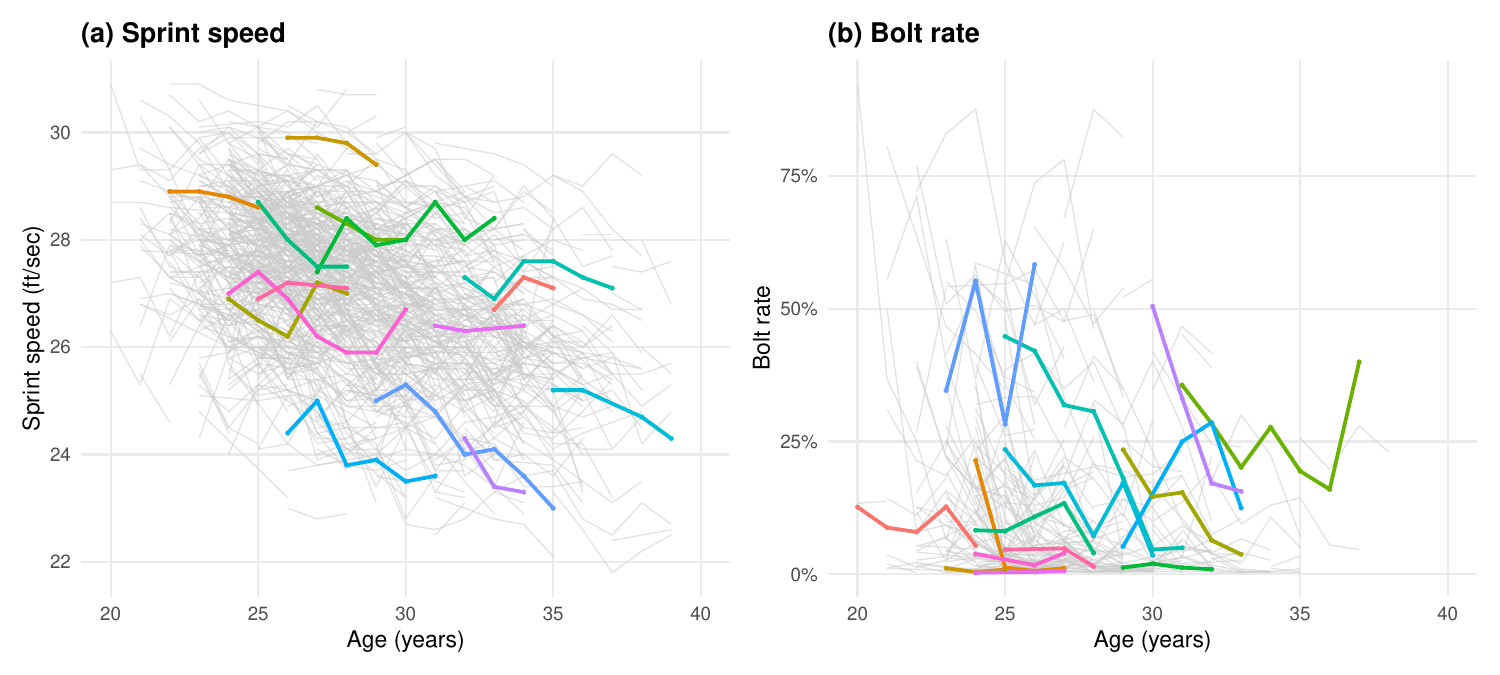}
\caption{Individual trajectories on the Stage-1 sample (same axes as Figure~\ref{fig:envelopes}): (a)~sprint speed ($n=1021$ athletes); (b)~bolt rate ($n=427$). Fifteen careers with at least three observations are highlighted.\label{fig:trajectories}}
\end{figure}

Figure~\ref{fig:envelopes} is the Stage~1 reference object for everything that follows. Panel~(a) shows age-conditional centiles of sprint speed under BCCGo: the solid C95 is nearly linear and decreasing over ages 20--40, so Stage~2 retains only $(\alpha_i,\delta_i)$ with $\gamma_i=0$ (Proposition~\ref{prop:linear}). Panel~(b) shows the corresponding Simplex centiles for bolt rate: the solid C95 is clearly curved, so Stage~2 uses the full $(\alpha_i,\gamma_i,\delta_i)$ geometry (Proposition~\ref{prop:curved}). Individual careers in Figure~\ref{fig:trajectories} and all later STAR fits are read relative to these solid ceilings, not relative to the mean.

\subsection{Insight 1: envelope geometry explains timing}\label{sec:geometry-insight}

Sprint timing is fixed at zero because the envelope is practically near-linear at the Stage~2 residual scale (Section~\ref{sec:ident}); bolt timing is identified ($\hat\omega_\gamma>0$) because the envelope is materially curved. The formal propositions alone are exact-linear and local; the applied contrast rests on the practical diagnostic in Section~\ref{sec:ident} and on Simulation Question~2. Table~\ref{tab:compare} summarizes the contrast and the Stage~2 fits that populate Functional Ageing Space in the next subsection.

\begin{table}[t]
\centering
\caption{Population summaries from single-MVN STAR-NLME (Laplace) relative to the Stage~1 GAMLSS C95.\label{tab:compare}}%
\begin{tabular*}{\linewidth}{@{\extracolsep\fill}lcc@{\extracolsep\fill}}
\toprule
Characteristic & Sprint speed & Bolt rate \\
\midrule
Players ($n_i\geq 3$) & 610 & 141 \\
Likelihood & Gaussian & Simplex \\
Envelope & GAMLSS C95 (BCCGo) & GAMLSS C95 (Simplex) \\
$\hat\mu_\alpha$ & $-0.94$ & $0.60$ \\
$\hat\mu_\gamma$ & 0 (fixed) & $0.23$ \\
$\hat\mu_\delta$ & $-0.20$ & $-0.46$ \\
$\hat\omega_\alpha$ & $2.13$ & $3.10$ \\
$\hat\omega_\delta$ & $0.39$ & $0.53$ \\
$\hat\rho_{\alpha\delta}$ & $-0.80$ & $-0.84$ \\
$\hat\sigma$ (residual) & $0.42$ & $3.88$ (Simplex) \\
Timing identifiable & No (linear envelope) & Yes \\
Level--tempo relationship & Strong negative & Strong negative \\
\botrule
\end{tabular*}
\begin{tablenotes}
\item Source: Primary Stage~2 Laplace fits (TMB); Stage~1 envelopes as in Figure~\ref{fig:envelopes}. Timing for sprint fixed per Table~\ref{tab:ident-geom}.
\end{tablenotes}
\end{table}

\begin{figure}[tbp]
\centering
\includegraphics[width=0.95\textwidth]{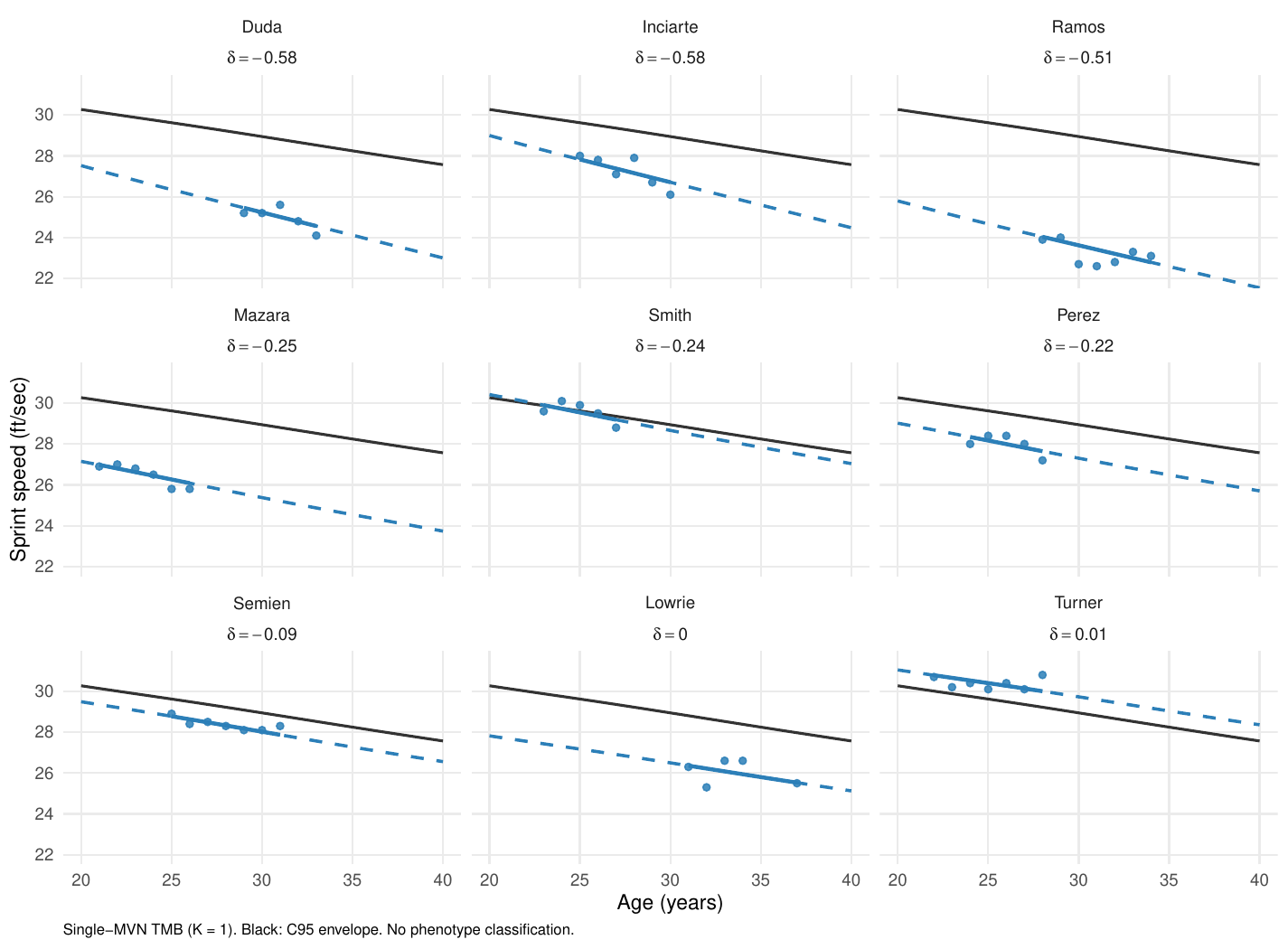}
\caption{Sprint speed: STAR fits from the single-MVN model for nine athletes spanning the $(\alpha,\delta)$ cloud. Black: envelope; blue: fitted trajectory.\label{fig:sprint-fits}}
\end{figure}

\begin{figure}[tbp]
\centering
\includegraphics[width=0.95\textwidth]{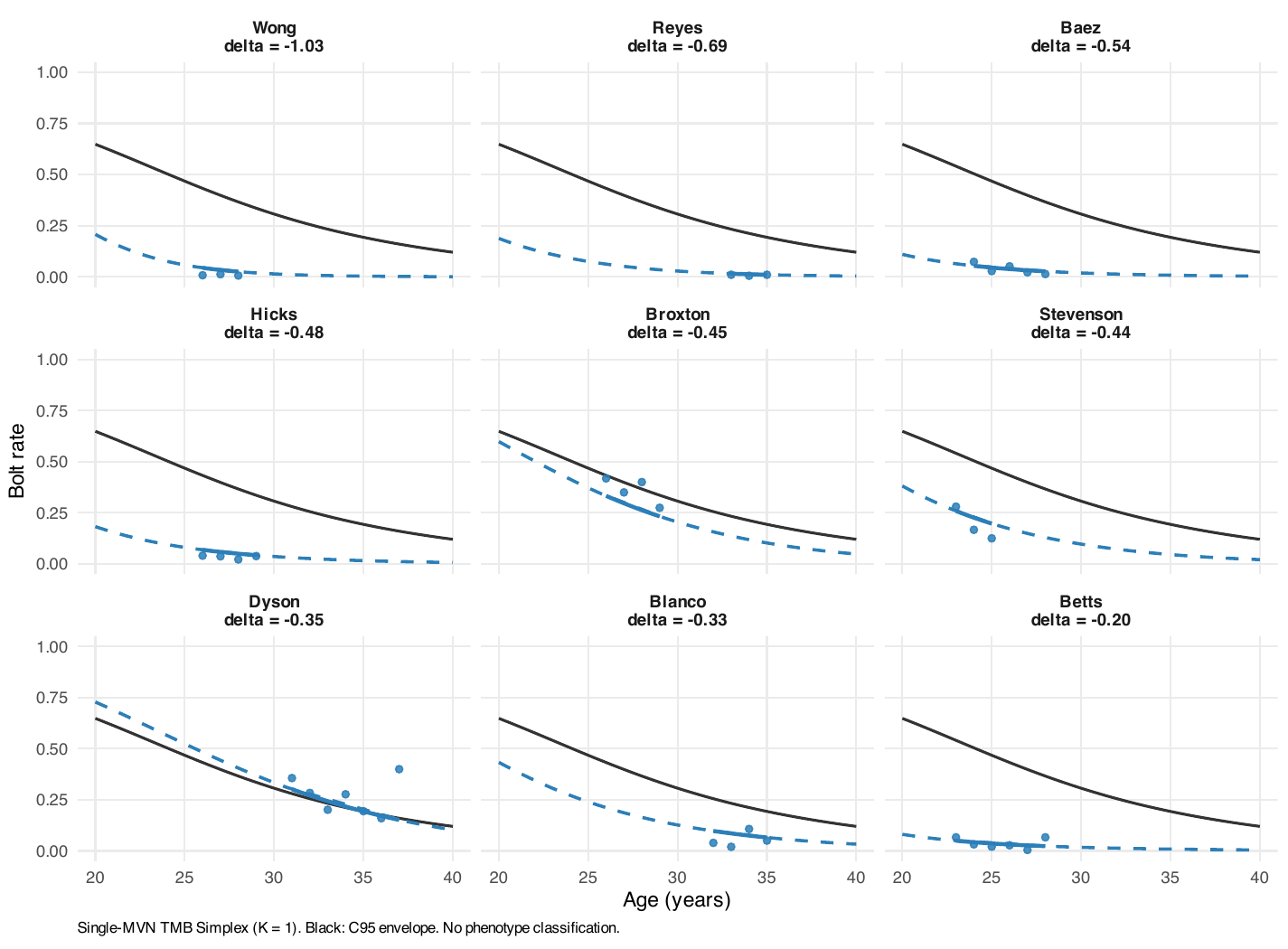}
\caption{Bolt rate: STAR fits from the single-MVN Simplex model for nine athletes. Black: C95; blue: fitted mean.\label{fig:bolt-fits}}
\end{figure}

Because the sprint envelope is approximately linear with slope $b\approx -0.14$~ft/sec per year, Proposition~\ref{prop:linear} requires $\gamma_i=0$ and Stage~2 retains $(\alpha_i,\delta_i)$ only. Bolt rate under the curved envelope and Simplex model~\eqref{eq:bolt} retains the full $(\alpha_i,\gamma_i,\delta_i)$ geometry; timing is nearly orthogonal to level and tempo ($\hat\rho_{\alpha\gamma}\approx\hat\rho_{\gamma\delta}\approx 0$). The Simplex dispersion $\hat\sigma_{\mathrm{S}}=3.88$ is not comparable in units to the Gaussian sprint residual $\hat\sigma_\varepsilon=0.42$. Illustrative individual fits appear in Figures~\ref{fig:sprint-fits}--\ref{fig:bolt-fits}.

\subsection{Insight 2: Functional Ageing Space}\label{sec:fas}

With the identifiable coordinates fixed by envelope geometry, each athlete is a point in Functional Ageing Space (FAS): level $\alpha$ (how close capacity sits to the envelope), tempo $\delta$ (how fast that capacity declines relative to the mean schedule), and---for bolt---timing $\gamma$. The statistical object of interest is no longer the trajectory itself, but these coordinates. The hierarchical association $\hat\rho_{\alpha\delta}$ in Table~\ref{tab:compare} is a parameter of $\Sigma$, not an artefact of correlating separate fits; higher-level athletes tend to have more negative tempo and thus steeper effective decline $b_i^*=b e^{-\delta_i}$ when the envelope is locally linear (Remark~\ref{rem:reparam}).

\begin{figure}[tbp]
\centering
\includegraphics[width=0.95\textwidth]{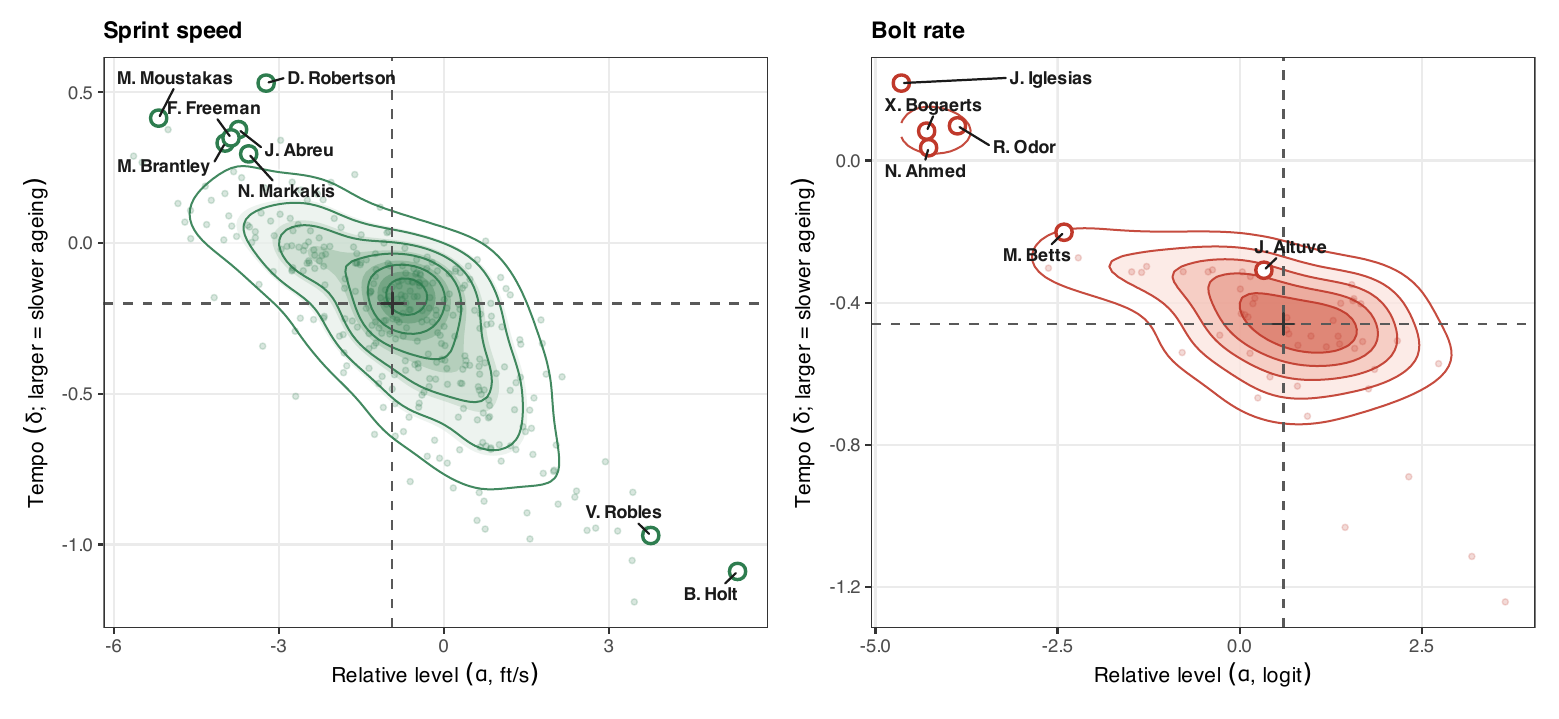}
\caption{Empirical distribution in Functional Ageing Space (FAS; design-restricted pool: $n_i\geq 5$, age span $\geq 4$). Contours show kernel density estimates of the empirical $(\alpha,\delta)$ distribution; dashed lines and cross: Stage~2 means $(\hat\mu_\alpha,\hat\mu_\delta)$; labeled points: careers in extreme regions of FAS. Relative to those means the quadrants are high/low level $\times$ slow/fast tempo. Sprint and bolt induce different cores and extremes; the descending ridge is $\hat\rho_{\alpha\delta}<0$.\label{fig:ageing-landscape}}
\end{figure}

Figure~\ref{fig:ageing-landscape} is the primary applied display: the empirical distribution in FAS for athletes with $n_i\geq 5$ and age span at least four years (sprint: $325$; bolt: $53$). Contours show kernel density estimates of the empirical $(\alpha,\delta)$ distribution of STAR BLUPs; dashed lines and the cross mark Stage~2 means $(\hat\mu_\alpha,\hat\mu_\delta)$; labeled points occupy extreme regions of FAS, not a hand-picked ``special player'' list. The density core is itself informative: the typical sprint career sits slightly below the envelope ($\hat\mu_\alpha=-0.94$) and ages slightly faster than the $\delta=0$ schedule ($\hat\mu_\delta=-0.20$). High-level/slow-tempo careers are rare under $\hat\rho_{\alpha\delta}=-0.80$; the descending ridge is that association as estimated inside $\Sigma$. We do not identify a mechanism---biological trade-off, mean reversion, differential exposure, or a feature of the parameterisation---without further analysis. Labeled slow agers (e.g.\ Moustakas, Freeman, Brantley, Robertson) lie outside the population core on the high-$\delta$ flank; labeled high-level fast agers (e.g.\ Holt, Robles) lie outside it on the low-$\delta$ flank. The bolt FAS cloud has a different core and different extremes (e.g.\ Altuve, Betts, Bogaerts, Odor), so the two metrics induce distinct FAS distributions.

\begin{figure}[tbp]
\centering
\includegraphics[width=0.72\textwidth]{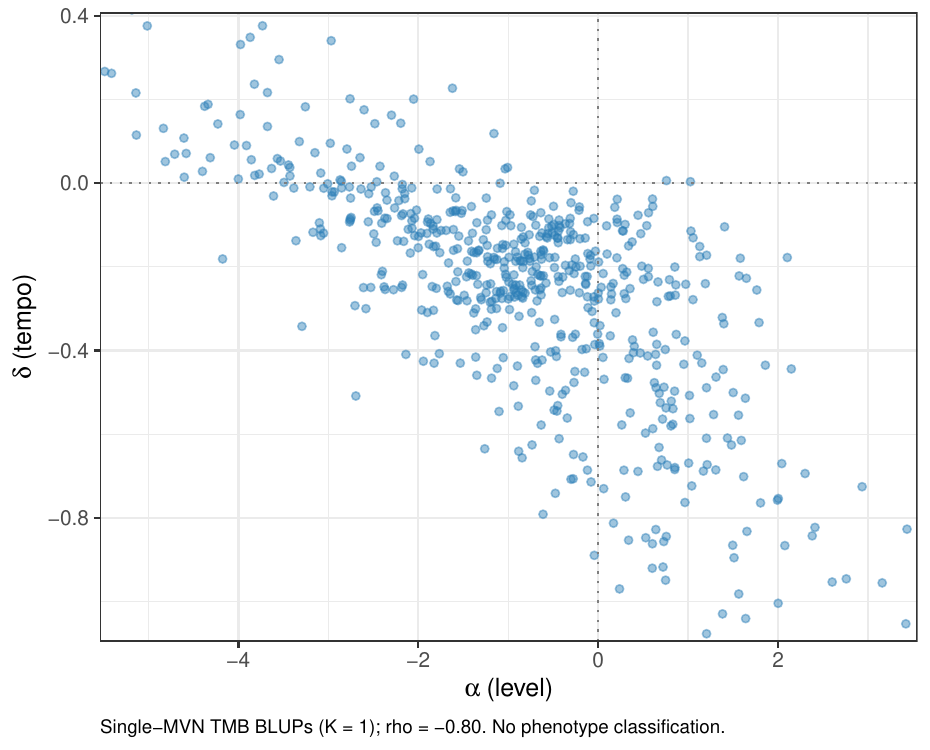}
\caption{Sprint speed: level versus tempo from single-MVN BLUPs ( $\hat\rho_{\alpha\delta}=-0.80$).}
\label{fig:alpha-delta-sprint}
\end{figure}

\begin{figure}[tbp]
\centering
\includegraphics[width=0.72\textwidth]{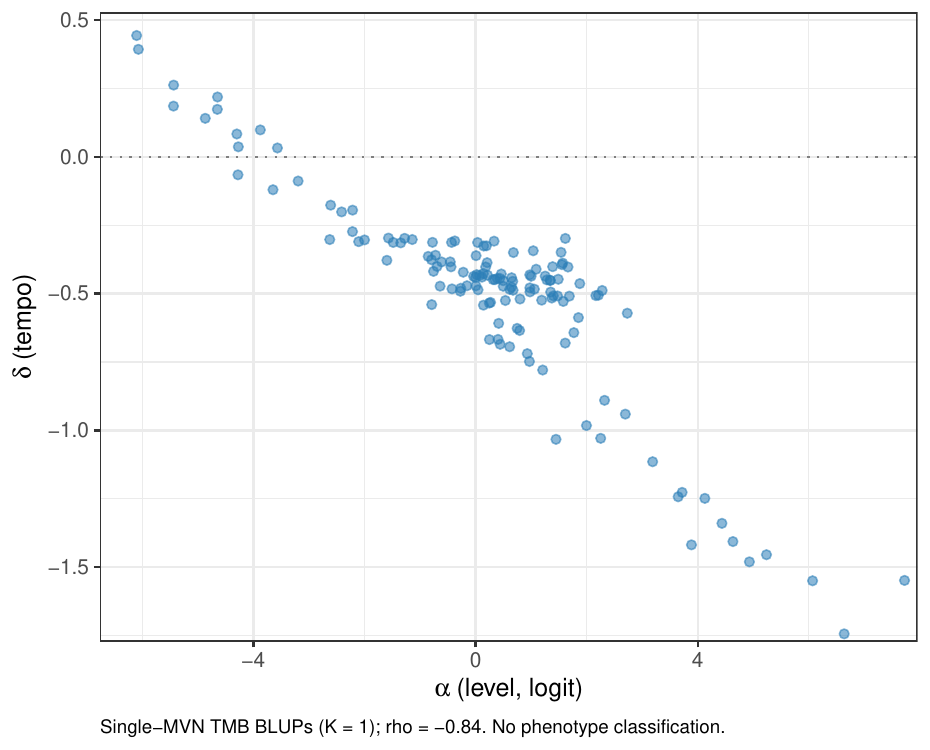}
\caption{Bolt rate: level (logit scale) versus tempo from single-MVN BLUPs ( $\hat\rho_{\alpha\delta}=-0.84$).}
\label{fig:alpha-delta-bolt}
\end{figure}

Figures~\ref{fig:alpha-delta-sprint}--\ref{fig:alpha-delta-bolt} show the unrestricted single-MVN BLUP clouds without the design restriction used for the landscape; they are diagnostic companions to Figure~\ref{fig:ageing-landscape}, not the primary reading.

\subsection{A derived tempo display: relative longevity}\label{sec:rli-display}

\begin{figure}[tbp]
\centering
\includegraphics[width=0.95\textwidth]{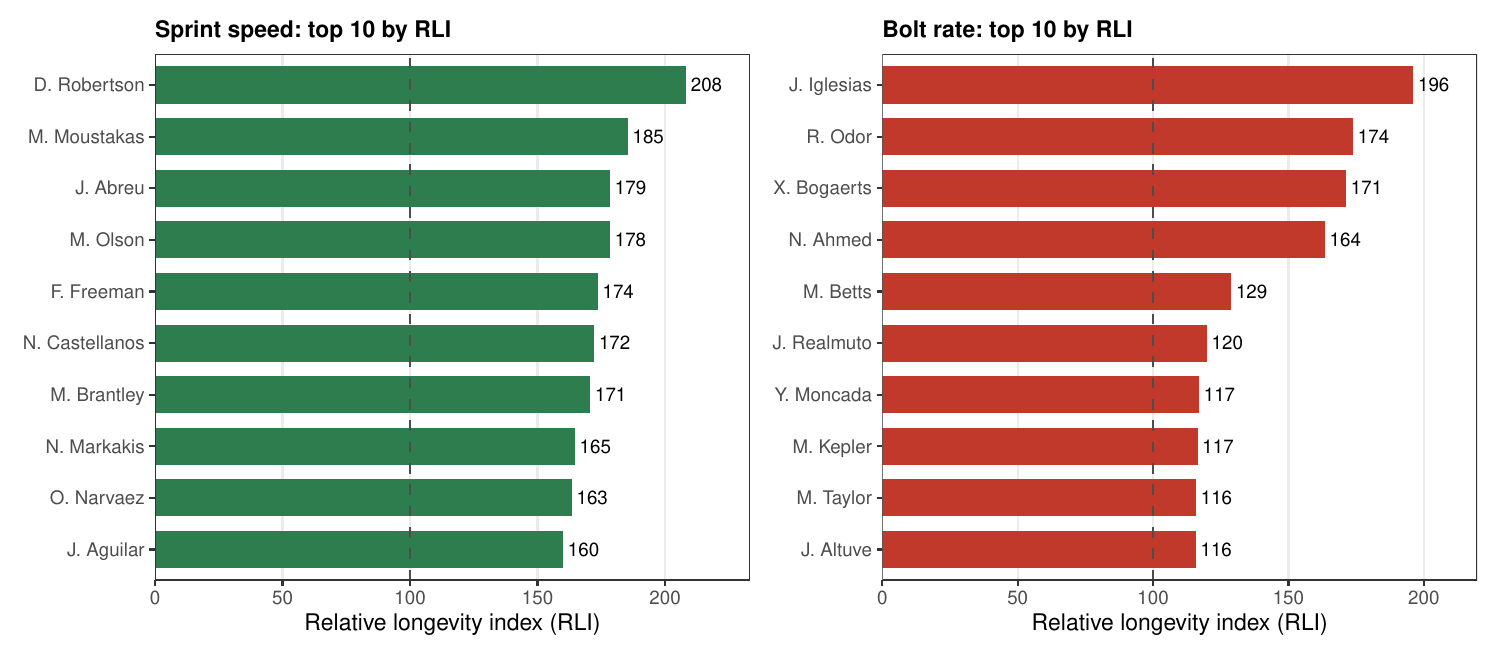}
\caption{Derived display: top ten by relative longevity index~\eqref{eq:rli} in the same design-restricted pool as Figure~\ref{fig:ageing-landscape}. RLI projects FAS tempo only. Dashed line: $\mathrm{RLI}=100$. Panels share no names.\label{fig:rli-top10}}
\end{figure}

Functional Ageing Space is the methodological object; a one-dimensional display of tempo alone is sometimes useful for communication. Define the \emph{relative longevity index}
\begin{equation}\label{eq:rli}
\mathrm{RLI}_i=100\cdot\exp(\delta_i-\hat\mu_\delta),
\end{equation}
so $\mathrm{RLI}=100$ is typical tempo, $120$ about $20\%$ slower decline, and $80$ about $20\%$ faster. RLI is a monotone projection of the FAS tempo coordinate onto a familiar scale---a derived summary, not a Stage~2 estimand and not the contribution of RACE/STAR. Figure~\ref{fig:rli-top10} ranks that projection in the same design-restricted pool used for Figure~\ref{fig:ageing-landscape}; the sprint and bolt top tens share no names, consistent with distinct FAS embeddings across metrics.

\subsection{Cross-metric reading of FAS}\label{sec:cross-metric}

A final check concerns the $141$ athletes observed on both metrics. Cross-metric concordance of $(\alpha,\delta)$ is only moderate for level and weak for tempo. Sprint speed measures near-maximal capacity; bolt rate measures success on aggressive baserunning attempts. They are related but distinct facets of running performance. What replicates is therefore the structural association $\rho_{\alpha\delta}<0$ \emph{under each metric's own envelope}, not a shared ranking of athletes across metrics. The performance--longevity message is about ageing relative to a chosen ceiling, not about a universal athlete ordering.

\section{Discussion}\label{sec:discussion}

The applied problem---sparse athletic careers read relative to peak attainable performance---forces a two-stage structure: estimate a performance envelope, then model individuals as geometric deviations from it within a mixed-effects hierarchy. RACE is that structure; STAR is the Stage~2 geometry; Functional Ageing Space is the primary statistical object in which athletes are represented. Envelope geometry determines which ageing parameters are identifiable; hierarchical estimation puts the level--tempo association inside $\Sigma$; RLI is only a derived projection of tempo.

For athletic ageing the demonstration yields two concrete insights. First (Section~\ref{sec:geometry-insight}), timing is not a meaningful individual parameter for near-linear sprint envelopes but is for curved bolt envelopes---a geometric fact, not a sample-size artefact. Second (Section~\ref{sec:fas}), Functional Ageing Space embeds careers as points in $(\alpha,\delta)$---with metric-specific cores and extremes and a strong negative level--tempo association inside $\Sigma$---of which RLI is only a derived tempo projection. Coordinates in FAS do not transfer cleanly across metrics (Section~\ref{sec:cross-metric}).

Several limitations follow from the same applied setting.

\paragraph{Identifiability scope.}
Proposition~\ref{prop:linear} is exact linearity; Proposition~\ref{prop:curved} is local near $\delta=0$. The applied near-linear/curved contrast in Table~\ref{tab:compare} therefore rests on the practical diagnostic in Section~\ref{sec:ident} and Simulation Question~2, not on a global theorem for the full $(\alpha,\gamma,\delta)$ domain.

\paragraph{Two-stage plug-in.}
Primary fits treat $\hat f$ as fixed. A player-cluster bootstrap that re-estimates the sprint envelope and Stage~2 ($B=199$ successful replicates; Supplementary Material, Section~E) gives median $\hat\rho_{\alpha\delta}\approx -0.70$ with percentile interval $(-0.83,-0.11)$ around the primary $-0.80$. The \emph{sign} is stable under envelope re-estimation; the interval is wide because a minority of resamples produce atypical high-centile envelopes that attenuate $|\hat\rho_{\alpha\delta}|$ toward zero, so the bootstrap median is a more representative summary of the bulk of the distribution than the right-tail extreme. We report the Stage~1--Stage~2 bootstrap for sprint only: bolt uses a weighted Simplex likelihood with three random effects on $n=141$ athletes, and cluster-resampling both the GAMLSS C95 and the 3-param Simplex NLME is substantially more fragile (non-convergence and unstable high-centile fits) than the Gaussian 2-param sprint case, so we do not claim a comparable bolt interval. Quantile/expectile plug-ins likewise leave $\hat\rho_{\alpha\delta}$ negative while reshaping the envelope; they do not displace GAMLSS as primary, and we do not claim geometry invariance across engines. Joint Stage~1--Stage~2 estimation remains a natural extension.

\paragraph{FAS versus index.}
Figure~\ref{fig:ageing-landscape} displays the empirical distribution in Functional Ageing Space. Section~\ref{sec:rli-display} introduces RLI only as a derived projection onto $\delta$; Figure~\ref{fig:rli-top10} ranks that projection. We do not interpret $\hat\rho_{\alpha\delta}<0$ as an identified biological trade-off.

\paragraph{Selection.}
Selection into continued play is unmodelled. One plausible (untested) mechanism---faster-tempo athletes exiting earlier---would attenuate $|\hat\rho_{\alpha\delta}|$, making $-0.80$ conservative rather than inflated; the reverse retention mechanism would push the other way. Primary results are single-MVN Laplace estimates.

We expect the same envelope-plus-hierarchy pattern wherever sparse series should be read relative to a capacity limit---growth, lung function, cognitive decline, rehabilitation, reliability testing, ecology \citep{lumeeker1993,meekerescobar1998}---with an analogous longevity display when tempo is identifiable. Transfer requires a scientifically meaningful envelope; we have not fit RACE/STAR outside athletic ageing. The contribution is the framework (envelope, STAR, hierarchical $\Sigma$, interpretable transforms), not a Statcast leaderboard.

Software for envelope estimation, STAR mixed models, and reproduction scripts is available at \url{https://github.com/idaejin/race-star-code} \citep{lee2026race}.

\section{Funding}\label{funding}

This work was partially funded by the SPHERES project (PID2023-153222OB-I00), awarded by MCIU/AEI/10.13039/501100011033 and by ``ERDF A way of making Europe'' (FEDER, UE).

\section{Acknowledgements}\label{acknowledgements}

MLB Statcast data were obtained via Baseball Savant. The author thanks Major League Baseball Advanced Media for making these data publicly available for research. AI-assisted writing tools were used for manuscript editing and drafting support; the author remains solely responsible for the scientific content, analyses, and interpretation.

\section{Disclosure statement}\label{disclosure-statement}

The author reports there are no competing interests to declare.

\section{Data Availability Statement}\label{data-availability-statement}

MLB Statcast data are publicly available through Baseball Savant. Analysis scripts, TMB templates, and reproduction materials are available at \url{https://github.com/idaejin/race-star-code} \citep{lee2026race}. The Supplementary Material documents envelope derivatives and identifiability (A), the Simplex PDF/CDF (B), Stage~2 hierarchical computation (C), computational notes (D), envelope diagnostics including P-spline edf, a Stage~1--Stage~2 bootstrap, and quantile/expectile $\hat\rho$ checks (E), and simulation design with summary tables for the three main-text questions (F). Stage~1 family choice (BCCGo, Simplex) is in the printed Appendix.

\begin{appendices}
\counterwithin{figure}{section}
\counterwithin{table}{section}
\renewcommand{\thefigure}{\thesection\arabic{figure}}
\renewcommand{\thetable}{\thesection\arabic{table}}

\section{Choice of GAMLSS families for Stage~1 envelopes}\label{app:families}\label{app:simplex}

Stage~1 uses the same smoother skeleton for both metrics---mean $\mathrm{pbm}(\mathrm{Age},\mathrm{mono}=\texttt{"down"})$, free scale $\mathrm{pb}(\mathrm{Age})$---and selects the observation family by GAIC among fits with a monotone-decreasing C95 (sports longevity) and near-nominal empirical coverage of that centile. The common skeleton is itself part of the GAMLSS case: one distributional engine, one monotone longevity constraint, and one scale smoother across metrics that would otherwise require incompatible Stage~1 recipes. Sprint (BCCGo) is treated first, then bolt (Simplex), matching the order in Section~\ref{sec:data}.

\subsection{Sprint speed: BCCGo}

Sprint speed is continuous and positive. We compared the Normal (NO) and BCCGo families \citep{rigbystasinopoulos2005,rigby2019}. Both yield monotone C95 envelopes. BCCGo has the better GAIC ($12\,112$ versus $12\,156$ for NO; $\Delta\mathrm{GAIC}\approx 44$) and empirical C95 coverage $95.0\%$ (NO: $96.0\%$). BCCGo allows mild skewness and a Box--Cox power on the positive support, which improves the upper-tail centile used as the RACE envelope (Figure~\ref{fig:envelopes}a). Stage~2 for sprint uses a Gaussian observation model on the STAR mean relative to this BCCGo C95 (the envelope geometry, not the Stage~1 skewness parameters, is what Stage~2 inherits).

\subsection{Bolt rate: Simplex}

\begin{figure}[tbp]
\centering
\includegraphics[width=0.92\textwidth]{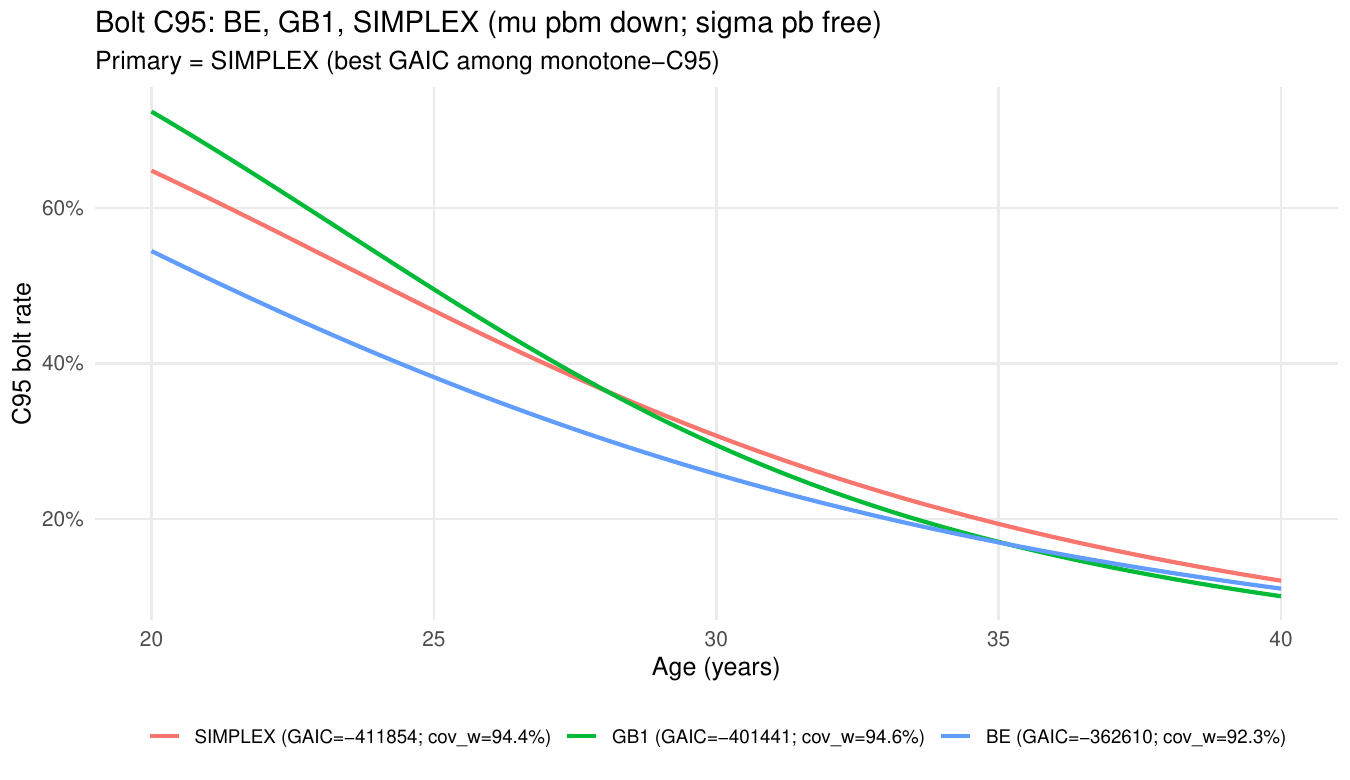}
\caption{Bolt-rate Stage~1 C95 under BE, GB1, and Simplex (common $\mathrm{pbm}$ mono-down mean and free $\mathrm{pb}$ scale). Legend: GAIC and opportunity-weighted C95 coverage. All three curves are monotone; Simplex minimizes GAIC and is primary.\label{fig:bolt-family}}
\end{figure}

Bolt rate lies in $(0,1)$ and is observed with opportunity weights equal to the number of competitive runs, so Stage~1 requires a unit-interval GAMLSS family. Under the common smoother we compared BE, GB1, and Simplex (SIMPLEX) \citep{barndorffjorgensen1991,rigbystasinopoulos2005,rigby2019}. Figure~\ref{fig:bolt-family} overlays the three C95 curves. Selection followed three ordered criteria: (i)~monotone-decreasing C95 (no late-age rebound); (ii)~minimize GAIC among eligible families; (iii)~prefer opportunity-weighted C95 coverage nearer $0.95$ as a soft tie-break.

All three families pass~(i). Criterion~(ii) separates them decisively: Simplex attains $\mathrm{GAIC}=-411\,854$, versus $-401\,441$ for GB1 ($\Delta\mathrm{GAIC}\approx 10\,413$) and $-362\,610$ for BE ($\Delta\mathrm{GAIC}\approx 49\,244$). Weighted coverage is near nominal for Simplex ($94.4\%$) and GB1 ($94.6\%$) and slightly low for BE ($92.3\%$), so~(iii) does not overturn the GAIC ranking. Visually, GB1 sits higher at young ages ($\widehat C_{95}(20)\approx 0.72$ versus $0.65$ for Simplex and $0.54$ for BE) and then drops more steeply; BE lies below the others through mid career; Simplex tracks an intermediate path ($0.65$ at~20, $0.31$ at~30, $0.12$ at~40) that is both the best-supported likelihood and the ceiling used in Figure~\ref{fig:envelopes}b. Stage~2 therefore uses the same Simplex observation model as Stage~1.

\end{appendices}

\label{supplementary-material}
\bigskip

\begin{center}
{\large\bf SUPPLEMENTARY MATERIAL}
\end{center}

\begin{unlist}
\item[\textbf{Title:~}] Supporting materials for ``Modelling Athletic Ageing Relative to an Estimated Performance Envelope'' by Dae-Jin Lee (full text appended after the references). Includes envelope derivatives and identifiability with a practical near-linearity diagnostic (A); Simplex PDF/CDF (B); Stage~2 hierarchical computation (C); computational notes (D); envelope diagnostics including P-spline edf, a Stage~1--Stage~2 bootstrap, and quantile/expectile $\hat\rho$ checks (E); and simulation design with summary tables for the three main-text questions (F). Stage~1 family choice (BCCGo, Simplex) is in the printed Appendix.
\item[\textbf{R and TMB code:~}]
 Analysis scripts, TMB templates, and reproduction materials at \url{https://github.com/idaejin/race-star-code} \citep{lee2026race}.
\item[\textbf{Processed data:~}]
 Deidentified athlete--season files for sprint speed and bolt rate used in Section~\ref{sec:application} are included in that repository.
\end{unlist}

\bibliographystyle{oup-abbrvnat}
\bibliography{bibliography}

\clearpage


\section*{Supplementary material}

# Supplementary Material for "Modelling Athletic Ageing Relative to an Estimated Performance Envelope"

Dae-Jin Lee
School of Science and Technology, IE University
Cardenal Zúñiga S/N, 40003 Segovia, Spain
Dae-Jin.Lee@ie.edu

**Abstract**

This supplement supports the main article. Section A derives envelope derivatives and proves the identifiability propositions, with a practical near-linearity diagnostic, a note on the local-only scope of Proposition 2, and interpretable STAR reparameterizations. Section B records the Simplex distribution used for bolt rate (PDF, CDF, and related quantities); Stage 1 family selection (BCCGo for sprint; Simplex for bolt) is in the main-text Appendix. Section C records Stage 2 optimization for the hierarchical single-MVN model. Section D gives computational notes. Section E records envelope diagnostics: the two-stage plug-in rationale, P-spline effective degrees of freedom, age-varying dispersion, a player-cluster bootstrap of Stage 1 into Stage 2 for sprint, and quantile/expectile Stage 1 checks that probe $\hat{\rho}_{\alpha\delta}$ without claiming geometry invariance. Section F records the simulation design and summary tables for the three main-text questions (correlation recovery, geometry and timing, envelope misspecification). Primary results in the article are single-MVN Laplace NLME estimates. Analysis code is available at https://github.com/idaejin/race-star-code.



## A Derivatives and Identifiability

Individual trajectories follow

$$y_i(x) = \alpha_i + f\left(\frac{x-\gamma_i}{\exp(\delta_i)}\right). \tag{1}$$

Let $u_i(x) = (x-\gamma_i)/\exp(\delta_i)$. Then

$$\frac{dy_i}{dx} = e^{-\delta_i} f'(u_i(x)), \qquad \frac{d^2y_i}{dx^2} = e^{-2\delta_i} f''(u_i(x)).$$

Thus $\delta_i > 0$ attenuates decline (greater longevity) and $\delta_i < 0$ amplifies it.

### A.1 Proof of Proposition 1 (linear envelopes)

If $f(x) = a + bx$ with $b \neq 0$, then

$$y_i(x) = \left(\alpha_i + a - b\gamma_i e^{-\delta_i}\right) + be^{-\delta_i}\,x. \tag{2}$$

The map $(\alpha_i, \gamma_i, \delta_i) \mapsto y_i(\cdot)$ depends on $(\alpha_i, \gamma_i)$ only through the linear combination $\alpha_i - b\gamma_i e^{-\delta_i}$. Hence for fixed $\delta_i$ the level sets of observationally equivalent $(\alpha_i, \gamma_i)$ are lines, and timing is

not separately recoverable from level. Given $b$, one recovers $\delta_i = -\log(d_i/b)$ from slope $d_i$. In applications with near-linear envelopes we set $\gamma_i = 0$ and estimate $(\alpha_i, \delta_i)$ only; the operational near-linearity check is given below.

The effective slope $b_i^* = be^{-\delta_i}$ is the annual rate of decline. For the empirical sprint-speed GAMLSS C95, $b \approx -0.14$ ft/sec per year (main text).

## A.2 Proof sketch of Proposition 2 (curved envelopes)

Write $m(x;\theta) = \alpha + f((x-\gamma)e^{-\delta})$ with $\theta = (\alpha, \gamma, \delta)$ and known $f \in C^2$. The Jacobian of $m(\cdot;\theta)$ with respect to $\theta$, evaluated at points $x_1, \ldots, x_n$ in an interval where $f'' \not\equiv 0$, has columns

$$\big(1,\ -e^{-\delta} f'(u(x)),\ -e^{-\delta}(x-\gamma) f'(u(x))\big).$$

When $f'$ is non-constant (equivalently $f'' \not\equiv 0$) and at least three distinct ages are observed, these columns are linearly independent in a neighbourhood of $\delta = 0$, so the structural mean map is locally injective and $(\alpha, \gamma, \delta)$ are locally identifiable. For a quadratic peak $f(x) = k - c(x - x^*)^2$, the individual peak is at $x_{\text{peak},i} = \gamma_i + x^* \exp(\delta_i)$ with curvature scaled by $e^{-2\delta_i}$, which separates timing from tempo. Bolt-rate envelopes under GAMLSS Simplex are monotone on the probability scale; on the logit scale they retain enough curvature for $(\alpha_i, \gamma_i, \delta_i)$ to be jointly identifiable in the application.

This argument is deliberately local. It does not deliver a global inverse over the full domain of $(\alpha, \gamma, \delta)$ used in the application tables. Simulation Question 2 in the main text bridges the gap: under exact linear truth, $\hat{\gamma}_i$ collapses; under quadratic and Gompertz truth with matched sparsity, $\gamma_i$ is recovered. We therefore treat Proposition 2 as a geometry diagnostic and the simulation as the finite-sample warrant for the applied claims.

## A.3 Practical near-linearity diagnostic

Proposition 1 is exact. Applications replace "exactly linear" by a noise-relative diagnostic on the fitted Stage 1 envelope $\hat{f}$ over ages $[20, 40]$. Let $r_{\text{lin}}$ be the residual root-mean-square of the ordinary linear fit of $\hat{f}$ on age, and let $\sigma$ be the Stage 2 residual scale. Stack the local Jacobian of $m(x; \alpha, \gamma, \delta)$ at $(\gamma, \delta) = (0, 0)$ on a sparse probe design $x \in \{22, 28, 34, 40\}$:

$$J(x) = \big(1,\ -f'(x),\ -x f'(x)\big).$$

A one-year timing shift then has mean-level size $\|J_{\cdot 2}\|_2$ on that design; the ratio $\|J_{\cdot 2}\|_2/\sigma$ is a signal-to-noise diagnostic for practical recovery of $\gamma$. For the primary sprint C95, $r_{\text{lin}} \approx 0.011$ ft/sec, linear $R^2 > 0.999$, and $\|J_{\cdot 2}\|_2/\hat{\sigma} \approx 0.3$ with $\hat{\sigma} \approx 0.42$: curvature is formally present but swamped by noise, so Stage 2 fixes $\gamma_i = 0$. For bolt rate on the probability scale the linear $R^2$ drops to $\approx 0.98$ and the C95 is visually curved; Stage 2 recovers $\hat{\tau}_\gamma > 0$. The diagnostic operationalises the paper's "near-linear" language; it is not a sharp threshold theorem. Analogous conditioning arguments for nonparametric components in structured additive regression (Kneib, 2013) motivate reporting such geometry checks alongside formal local-identifiability statements.

## A.4 Interpretable reparameterizations

Equation (1) is the structural mean used for estimation. For reporting, one-to-one transforms can be clearer without changing the model (main-text Remark on interpretable reparameterization). Under a locally linear envelope,

$$\mu_i(x) = L_i + S_i(x - x_0), \qquad L_i = f(x_0) + \alpha_i, \qquad S_i = b\, e^{-\delta_i} = b_i^*,$$

so $L_i$ is level at a reference age and $S_i$ is absolute annual decline. Equivalently $\lambda_i = e^{\delta_i}$ reads as calendar years per unit of envelope age. For a curved envelope with peak $x^\star$ of $f$, the individual peak $x_{\text{peak},i} = \gamma_i + x^\star e^{\delta_i}$ is often clearer than raw $(\gamma_i, \delta_i)$. These are display transforms of STAR parameters, not alternative Stage 2 geometries.

## B The Simplex Distribution

Bolt rate $Y \in (0,1)$ is modeled with the univariate Simplex distribution of Barndorff-Nielsen and Jørgensen (1991), as implemented in `gamlss.dist` (Rigby and Stasinopoulos, 2005; Rigby et al., 2019). The family is a proper dispersion model on the unit interval with mean parameter $\mu \in (0,1)$ and dispersion/scale $\sigma > 0$. Selection of Stage 1 GAMLSS families (BCCGo for sprint; Simplex for bolt) is documented in the main-text Appendix. This section records the Simplex PDF/CDF used for bolt rate in both stages.

### B.1 Probability density function

The PDF used in Stage 1 (GAMLSS) and Stage 2 (TMB) is

$$f(y \mid \mu, \sigma) = \left(2\pi\sigma^2\, y^3(1-y)^3\right)^{-1/2} \exp\!\left(-\frac{1}{2\sigma^2}\,\frac{(y-\mu)^2}{y(1-y)\,\mu^2(1-\mu)^2}\right), \qquad 0 < y < 1. \tag{3}$$

Equivalently, writing the unit deviance

$$d(y;\mu) = \frac{(y-\mu)^2}{y(1-y)\,\mu^2(1-\mu)^2},$$

the density is the dispersion-model form

$$f(y \mid \mu, \sigma) = \left(2\pi\sigma^2\,\{y(1-y)\}^3\right)^{-1/2} \exp\!\left(-d(y;\mu)/(2\sigma^2)\right).$$

The log-density implemented in `star_bolt_simplex.cpp` matches (3) up to floating-point clipping of $y$ and $\mu$ away from $\{0,1\}$. Default GAMLSS links are $\operatorname{logit}(\mu)$ and $\log(\sigma)$.

### B.2 Mean, variance function, and moments

Under (3), $\mathrm{E}(Y) = \mu$. The unit variance function of the dispersion model is

$$V(\mu) = \mu^3(1-\mu)^3,$$

so that small-dispersion asymptotics give $\operatorname{Var}(Y) \approx \sigma^2 V(\mu) = \sigma^2\mu^3(1-\mu)^3$. There is no closed-form expression for higher moments beyond this asymptotic variance; exact moments are obtained by numerical integration against (3) when needed.

### B.3 CDF, quantile function, and simulation

The CDF has no closed form. In `gamlss.dist` it is obtained by numerical quadrature of the PDF,

$$F(q \mid \mu, \sigma) = \int_0^q f(y \mid \mu, \sigma)\, dy, \qquad 0 < q < 1, \tag{4}$$

with the conventions $F(q) = 0$ for $q \le 0$ and (by truncation of the integrand support) the usual unit-interval boundaries handled by the package. The quantile function $F^{-1}(p \mid \mu, \sigma)$ is the numerical inverse of (4); Stage 1 bolt envelopes extract the age-conditional 95th centile via this inverse (through `centiles.pred` / `qSIMPLEX`). Random draws use the inverse-CDF transform $U \sim \mathrm{Unif}(0,1)$, $Y = F^{-1}(U \mid \mu, \sigma)$.

### B.4 Role in RACE/STAR

At Stage 1, $Y_{ij} \mid x_{ij} \sim \mathrm{SIMPLEX}(\mu(x_{ij}), \sigma(x_{ij}))$ with opportunity weights equal to competitive runs; the RACE envelope is the fitted 0.95 centile of this conditional law. At Stage 2 the same PDF enters the player-level weighted log-likelihood with $\operatorname{logit}(\mu_{ij}) = \alpha_i + \hat{f}\big((x_{ij} - \gamma_i)/\exp(\delta_i)\big)$ and a global $\sigma$ (TMB), matching the observation model of the Stage 1 family choice.

# C Stage 2 Optimization: TMB STAR-NLME

Stage 2 uses the same STAR mean relative to the Stage 1 envelope $\hat{f}$ under two random-effect laws that answer different applied questions (main-text Estimation section).

## C.1 Envelope interpolant

The Stage 1 envelope is a numeric grid, not an analytic function. After fitting GAMLSS we evaluate

$$\hat{f}(x_g) = \widehat{F}^{-1}_{Y|x_g}(0.95), \qquad g = 1, \ldots, G,$$

on a padded age grid (e.g. $G = 301$ points on $[15, 45]$) via `centiles.pred`. For bolt rate the Stage 2 grid is stored on the logit scale so that STAR operates as a linear predictor before the inverse logit. Stage 2 defines

$$\hat{f}(u) = \texttt{approxfun}(\{x_g\}, \{\hat{f}(x_g)\})(u)$$

with linear interpolation inside the grid and edge-slope extrapolation outside. Transformed ages $u_{ij} = (x_{ij} - \gamma_i)/\exp(\delta_i)$ are evaluated at arbitrary points inside the TMB C++ template.

## C.2 Hierarchical single-MVN model

The primary Stage 2 fit places a single multivariate normal on the random effects. Sprint speed uses a Gaussian observation model with $(\alpha_i, \delta_i)^\top \sim \text{MVN}(\mu, \Sigma)$, $\Sigma$ parameterized by $(\tau_\alpha, \tau_\delta, \rho_{\alpha\delta})$. Bolt rate uses a weighted Simplex likelihood with $(\alpha_i, \gamma_i, \delta_i)^\top \sim \text{MVN}(\mu, \Sigma)$ (unstructured correlation factor). In both cases the integrated likelihood is approximated by Laplace's method with automatic differentiation (Kristensen et al., 2016); optimization uses `nlminb`. Population parameters $(\mu, \Sigma, \sigma)$ maximize the Laplace approximate marginal likelihood; athlete summaries are BLUPs shrunk toward the single MVN.

This model is the estimand for the *population* level–tempo association: $\rho_{\alpha\delta}$ is a parameter of $\Sigma$. It supplies Insights 1–2 in the main text (sign and magnitude of $\rho_{\alpha\delta}$; which of $\alpha, \gamma, \delta$ are recovered under each envelope) and the coordinates that define Functional Ageing Space. Implementation: `star_sprint.cpp`, `star_bolt_simplex.cpp`, and the fitting scripts in the accompanying repository (Lee, 2026) (`https://github.com/idaejin/race-star-code`).

# D Computational Details

Analyses used R 4.4+ with `gamlss`, `TMB`, `dplyr/readr`, and `ggplot2`. Code and reproduction scripts are at `https://github.com/idaejin/race-star-code` (Lee, 2026). Envelope estimation is seconds. Compiling the TMB templates (`star_sprint.cpp`, `star_bolt_simplex.cpp`) is a one-time cost of about a minute. Fitting the global sprint NLME ($n = 610$) and bolt Simplex NLME ($n = 141$) is typically under a minute each on an Apple M1 laptop after compilation. Simulation factorials are several hours depending on the number of Monte Carlo replicates. The player-cluster Stage 1–Stage 2 bootstrap (script `11_bootstrap_envelope_star.R`) is reported for sprint only ($\approx 10$ minutes for $B = 200$ on the same hardware when it completes without early optimizer stalls); an analogous bolt bootstrap is not reported because the weighted Simplex, three-random-effect fit is substantially more fragile under cluster resampling.

# E Additional Envelope Diagnostics

## E.1 Why a two-stage plug-in?

Stage 1 estimates a *population* age-conditional high centile from pooled athlete-seasons—a growth-chart object in the sense of LMS/GAMLSS (Cole and Green, 1992; Rigby and Stasinopou-

Table 1: Stage 1 P-spline effective degrees of freedom and penalty weights for the primary envelopes.

| Metric | Family | $\mu$ edf (pbm) | $\sigma$ edf (pb) | $\lambda_\mu$ |
|---|---|---|---|---|
| Sprint | BCCGo | 2.71 | 2.01 | $9.6 \times 10^6$ |
| Bolt | Simplex | 2.04 | 2.09 | $1.0 \times 10^7$ |

Source: Primary GAMLSS fits (`gamlss`); smoother skeleton as in the main-text Appendix. edf from the fitted P-spline penalties.

los, 2005). Stage 2 then models sparse individual careers as low-dimensional geometric transforms of that chart. The split is deliberate: the scientific target of Stage 1 is the attainable ceiling at age $x$, not a random-effects mean entangled with individual $(\alpha_i, \gamma_i, \delta_i)$. A single joint additive predictor that folds STAR random effects into the same fit as the centile would redefine $f$ and change the estimand. Growth-reference practice is likewise two-stage (fit the chart, then assess individuals relative to it). The cost is that primary Stage 2 fits treat $\hat{f}$ as fixed; we quantify the resulting sensitivity by a player-cluster bootstrap below rather than relying on a delta-method approximation to a joint Hessian that the present pipeline does not deliver.

## E.2 P-spline effective degrees of freedom

Both primary Stage 1 fits use the same smoother skeleton: monotone-decreasing P-splines for location, pbm(Age, mono = `"down"`), and unrestricted P-splines for scale, pb(Age). Smoothing parameters are chosen by the local maximum-likelihood / GAIC criterion internal to `gamlss`. Table 1 reports the effective degrees of freedom (edf) and the fitted penalty weights $\lambda$ for the primary models. Location edf is low for both metrics (sprint $\approx 2.7$; bolt $\approx 2.0$), so the *mean* smoother is nearly linear under the monotone constraint. That is consistent with the near-linear sprint C95 in the main text. For bolt rate, the RACE envelope is the Simplex age-conditional 95th centile, not $\mu(x)$ itself: age-varying scale and the nonlinear quantile map can still produce a clearly curved C95 even when the location edf is near two. Identifiability propositions in the main text therefore concern the geometry of $f = \widehat{C}_{95}$, not the edf of $\mu$ alone.

## E.3 Age-varying dispersion (sprint)

Under a Normal GAMLSS fit for sprint speed (same smoother skeleton), $\hat{\sigma}(25) \approx 1.29$, $\hat{\sigma}(30) \approx 1.39$, and $\hat{\sigma}(35) \approx 1.49$ ft/sec, indicating increasing conditional dispersion with age.

## E.4 Player-cluster bootstrap of Stage 1 plug-in (sprint)

To quantify sensitivity of Stage 2 summaries to Stage 1 uncertainty we resample athletes with replacement (preserving within-athlete seasons), refit the GAMLSS C95, and refit the Laplace STAR-NLME on the new envelope (script `11_bootstrap_envelope_star.R`; $B = 200$, 199 successful replicates). Figure 1 shows pointwise bootstrap bands for the sprint envelope; Figure 2 shows the bootstrap distribution of $\hat{\rho}_{\alpha\delta}$ and $\hat{\mu}_\delta$. The primary point estimate $\hat{\rho}_{\alpha\delta} = -0.80$ sits in the left tail of a still strongly negative bootstrap distribution (median $\approx -0.70$; percentile interval $(-0.83, -0.11)$). The upper endpoint near zero is driven by a minority of resamples that yield atypical ceilings; the median better summarizes the bulk of the distribution. The sign of the level–tempo association is therefore stable under envelope re-estimation, while the magnitude is sensitive to which athletes define the ceiling—as expected for a high-centile plug-in. We restrict this Stage 1–Stage 2 bootstrap to sprint speed. Bolt rate combines a weighted Simplex Stage 1/Stage 2 pipeline with three random effects on a smaller athlete panel ($n = 141$); cluster-resampling both layers is substantially more prone to non-convergence and unstable

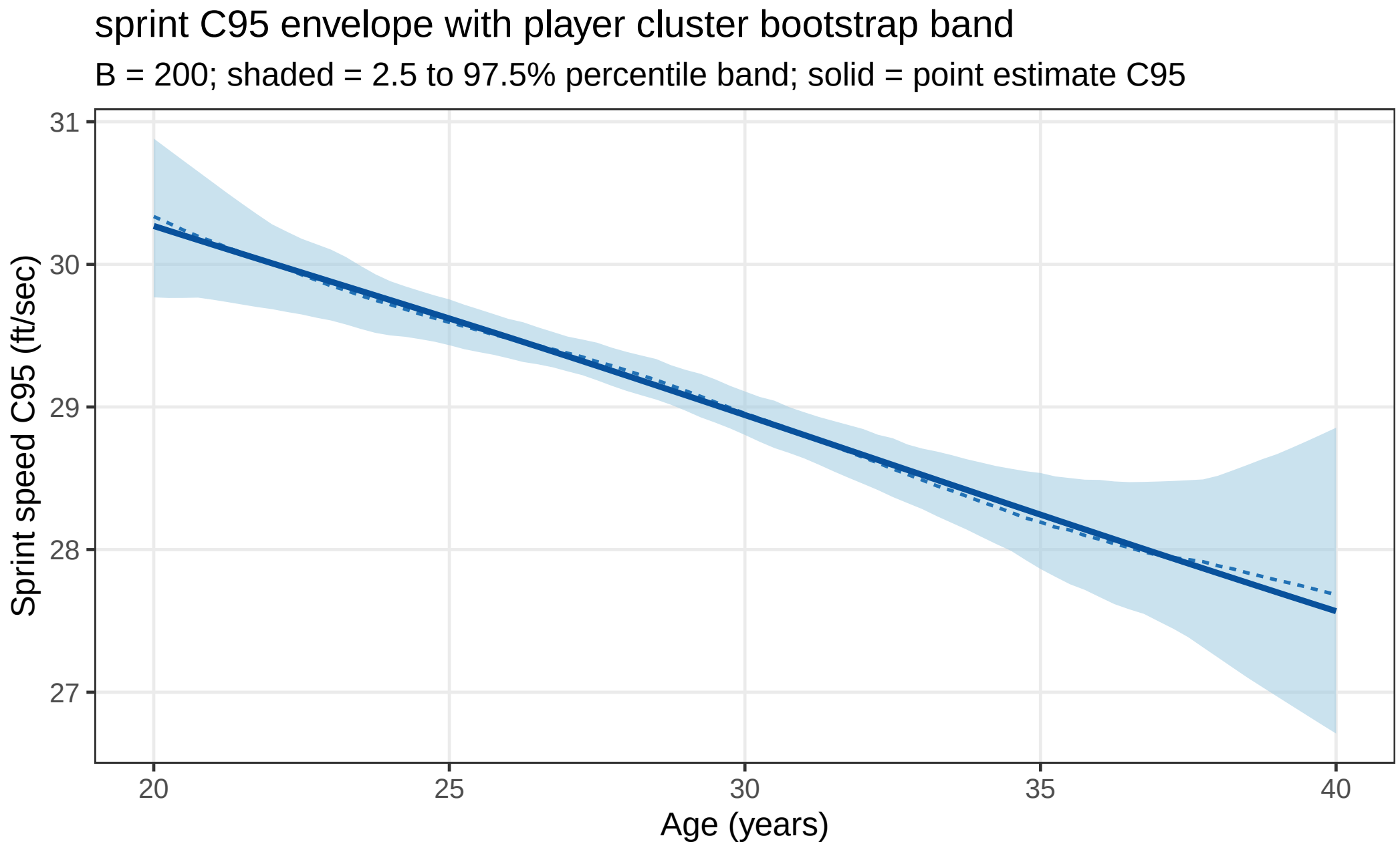


Figure 1: Sprint speed: player-cluster bootstrap bands for the Stage 1 C95 envelope ($B = 199$ successful replicates) with the primary GAMLSS fit overlaid.

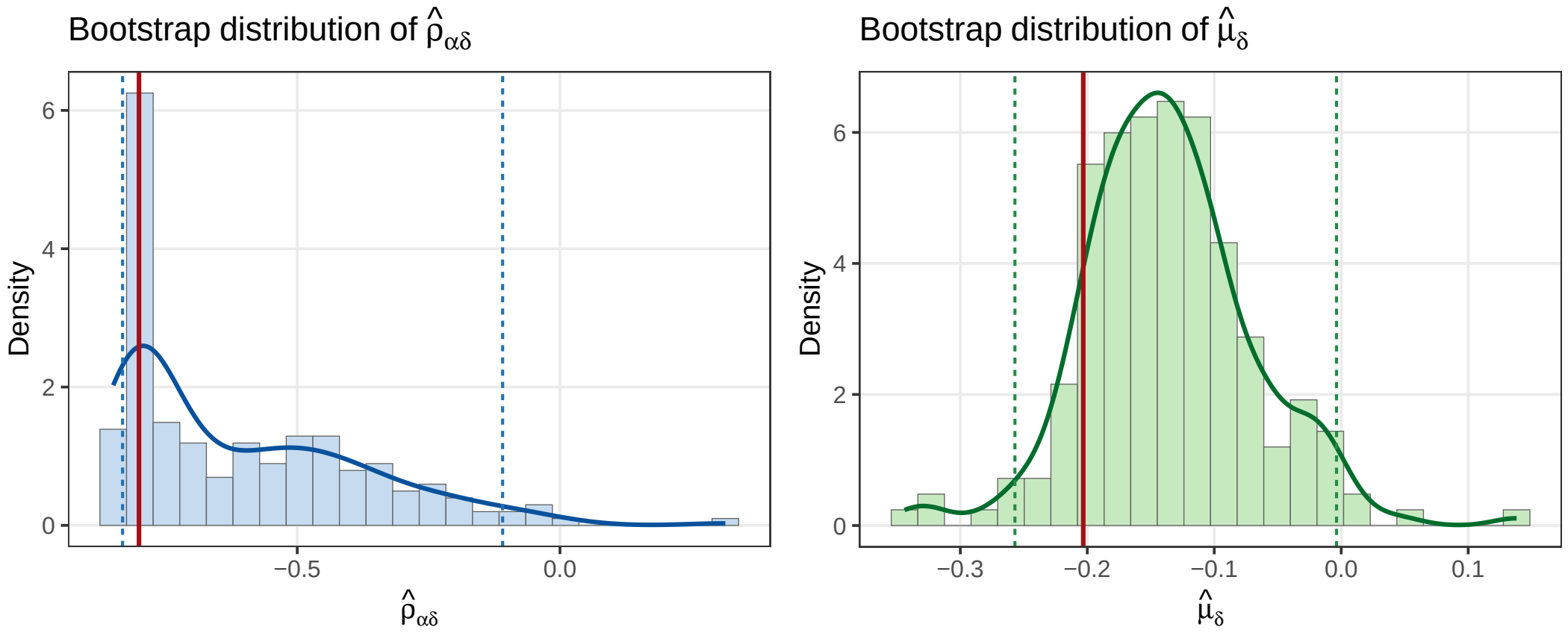


Figure 2: Sprint speed: bootstrap distributions of $\hat{\rho}_{\alpha\delta}$ and $\hat{\mu}_\delta$ from refitting Stage 1 and Stage 2 on player-cluster resamples ($B = 199$). Vertical lines mark the primary point estimates.

upper-centile fits than the Gaussian two-parameter sprint case, so a comparable bolt interval is not reported. Full joint Stage 1–Stage 2 estimation remains a natural extension; the sprint bootstrap already answers the immediate applied question of whether the reported association is an artefact of a single fixed $\hat{f}$.

## E.5 Stage 1 engine checks (quantile and expectile)

The primary envelope is the GAMLSS age-conditional 95th centile: a growth-chart object estimated by distributional regression, with age-varying scale, metric-matched families, an in-model monotone mean, and a coherent multi-centile fan (main text, Section 2.1). Direct quantile regression targets the same estimand $F^{-1}_{Y|x}(\tau)$ without a full conditional distribution (Koenker, 2005); expectiles are a different functional (Newey and Powell, 1987) and enter only as a stress probe. We fit B-spline quantile regression and asymmetric least-squares expectiles at $\tau = 0.95$

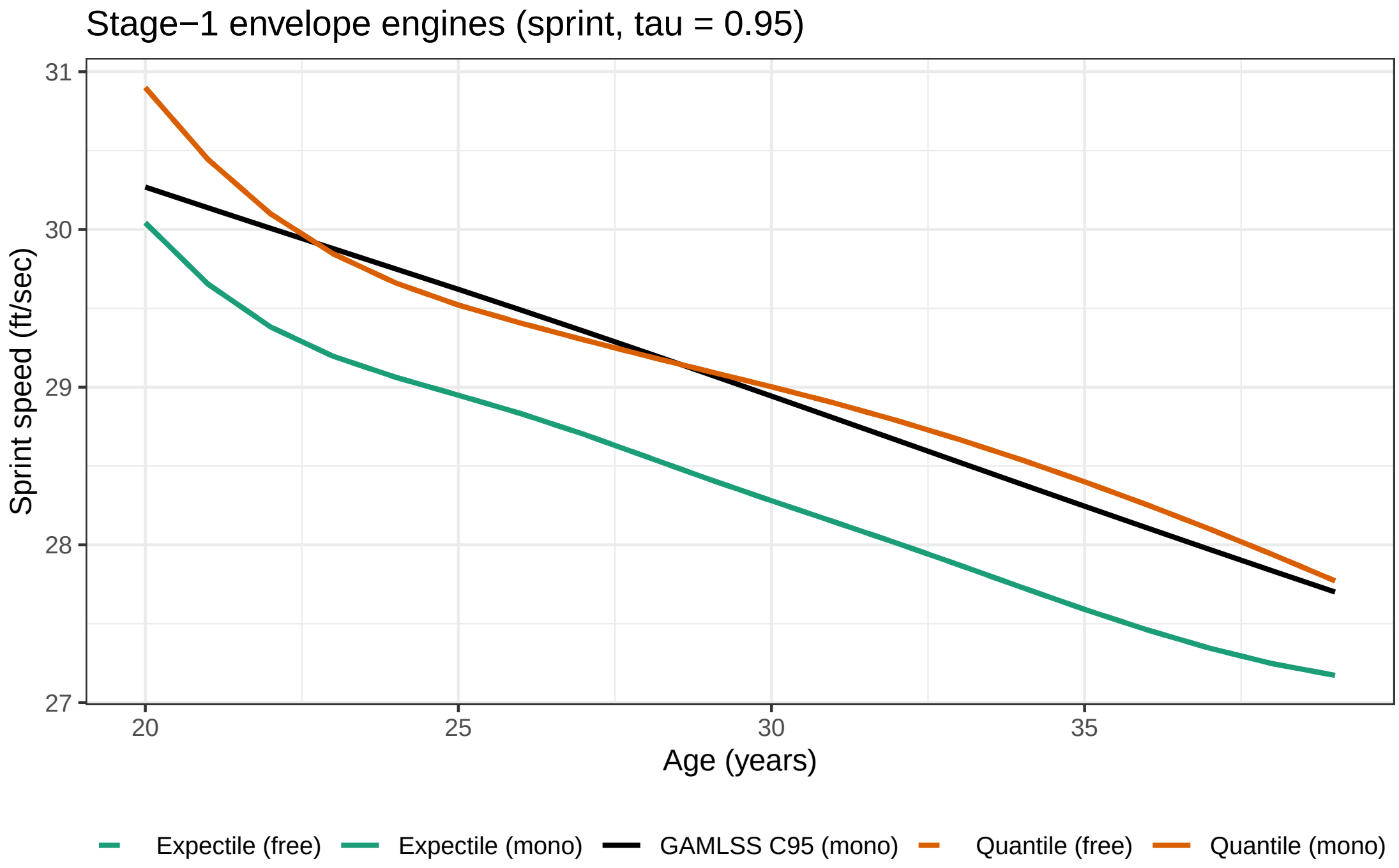


Figure 3: Sprint Stage 1 curves at $\tau = 0.95$: primary GAMLSS C95 versus quantile and expectile fits (free) and post-hoc monotone SCAM projections.

(bolt: opportunity-weighted), report the unconstrained curves, and—as a soft longevity regularizer only—optionally project each fitted curve onto a monotone-decreasing path from a shape constrained additive model (SCAM; Pya and Wood, 2015) with basis `bs='mpd'`. That SCAM step is a post-hoc shape projection on the estimated curve (unique ages), not a joint monotone quantile/expectile likelihood and not comparable to GAMLSS pbm(mono = `"down"`) inside the observation model. The contrast underscores why GAMLSS remains primary: competitors either omit the conditional law (quantile) or estimate the wrong functional (expectile), and neither embeds sports-longevity monotonicity in the Stage 1 likelihood. Script: `15_envelope_engine_robustness.R`.

Figures 3–4 overlay free and projected curves on the GAMLSS C95. Unconstrained quantile fits are variable in the upper tail; monotone projections reshape them and need not track GAMLSS—exactly the instability one expects when a thin upper tail is estimated without a distributional scale model. Expectile paths are smoother and closer to GAMLSS for sprint, but an expectile is still not a C95. Table 2 reports Stage 2 Laplace STAR-NLME fits that plug in the monotone projections (sprint with $\gamma$ fixed as in the primary analysis). The level–tempo association stays strongly negative (sprint $\hat{\rho}_{\alpha\delta} \in \{-0.79, -0.79, -0.75\}$; bolt $\in \{-0.77, -0.70, -0.92\}$). Envelope *geometry*, however, is partly engine-dependent: linear $R^2$ of the Stage 1 curve over ages 20–40 is $> 0.999$ for GAMLSS sprint C95, about 0.988 for the expectile projection, and about 0.966 for the quantile projection. We therefore treat GAMLSS as the primary growth-chart ceiling that justifies the linear/curved Stage 2 design, claim robustness of $\hat{\rho}_{\alpha\delta}$ under alternative plug-ins, and do *not* claim invariance of envelope shape across engines.

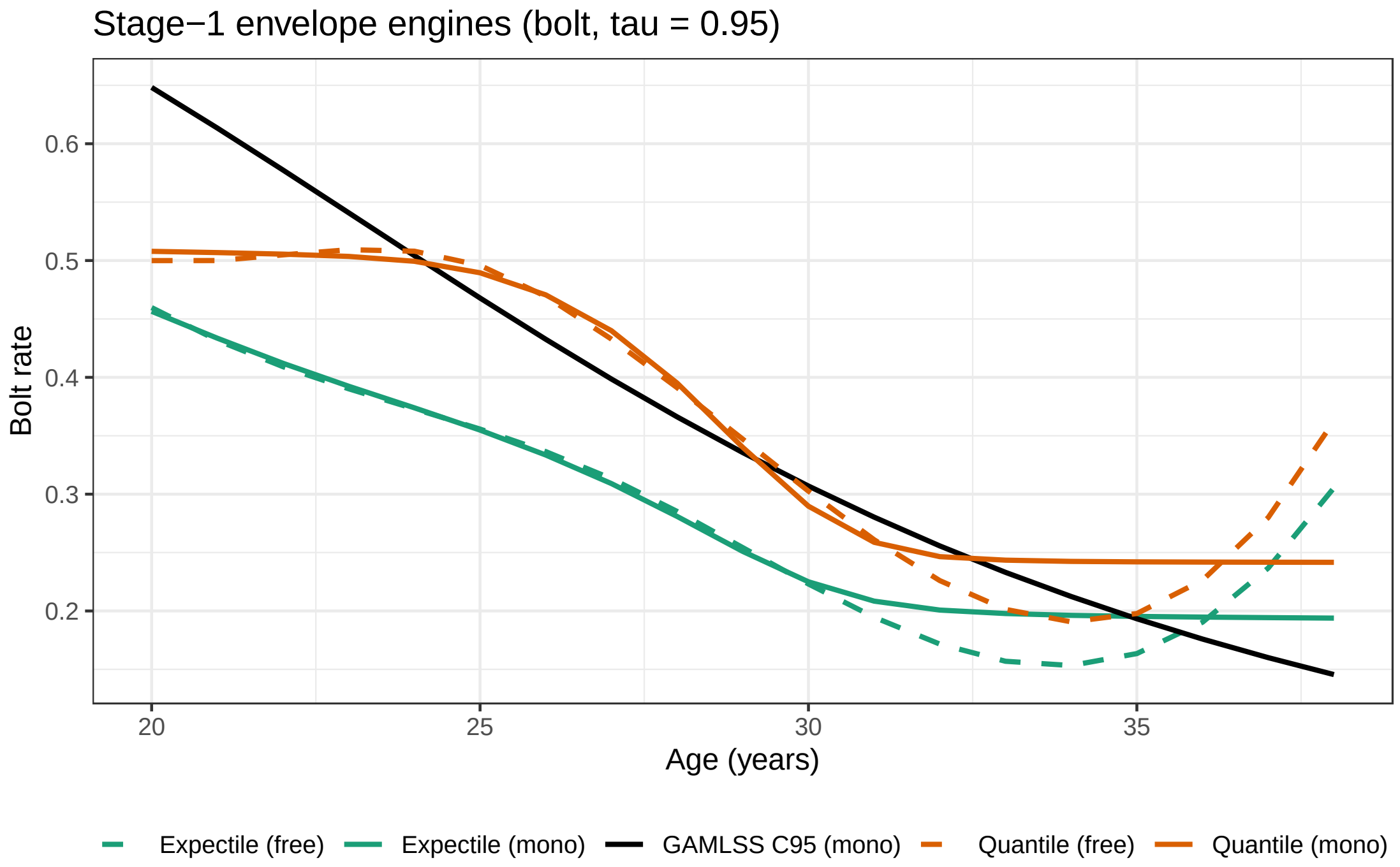


Figure 4: Bolt Stage 1 curves at $\tau = 0.95$: primary GAMLSS C95 versus quantile and expectile fits (free) and post-hoc monotone SCAM projections.

Table 2: Stage 2 Laplace STAR-NLME $\hat{\rho}_{\alpha\delta}$ under the primary GAMLSS C95 and under monotone SCAM projections of quantile/expectile curves (sensitivity plug-ins).

| Metric | Stage 1 plug-in | $\hat{\rho}_{\alpha\delta}$ | $\hat{\mu}_{\delta}$ |
|---|---|---|---|
| Sprint | GAMLSS C95 (primary) | −0.79 | −0.21 |
| Sprint | Quantile + SCAM projection | −0.79 | −0.24 |
| Sprint | Expectile + SCAM projection | −0.75 | −0.11 |
| Bolt | GAMLSS C95 (primary) | −0.77 | −0.38 |
| Bolt | Quantile + SCAM projection | −0.70 | −0.74 |
| Bolt | Expectile + SCAM projection | −0.92 | −1.01 |

Source: Script `15_envelope_engine_robustness.R`. Quantile (Koenker, 2005); expectile (Newey and Powell, 1987); SCAM projection (Pya and Wood, 2015). Primary GAMLSS C95 remains the scientific reference; SCAM rows are post-hoc monotone projections, not peer Stage 1 engines.

# F Simulation details

## F.1 Design

We simulate from known envelopes—linear, quadratic, and Gompertz—with individual parameters drawn from an MVN with known $\rho_{\alpha\delta} \in \{-0.8, -0.4, 0, 0.4\}$. Sparsity is set to $n_i \in \{3, 5, 8\}$ observations per individual and residual scales matched to the application. For each scenario we compare: (a) Laplace STAR-NLME; (b) separate per-athlete STAR fits with post-hoc correlation of $(\hat{\alpha}_i, \hat{\delta}_i)$; (c) a random-intercept-and-slope Gaussian mixed model on the raw scale. Monte Carlo size is $B = 50$ replicates per cell (1,800 cells in the full factorial). Tables 3–5 summarize the three main-text questions; replicate-level files ship with the reproduction archive.

Table 3: Simulation Question 1 (linear envelope): Monte Carlo mean $\hat{\rho}_{\alpha\delta}$ from hierarchical STAR-NLME versus post-hoc correlation of separate fits ($B = 50$).

| True $\rho_{\alpha\delta}$ | $n_i$ | Hierarchical | Separate |
|---|---|---|---|
| −0.8 | 3 | −0.80 | −0.83 |
| −0.8 | 5 | −0.79 | −0.82 |
| −0.8 | 8 | −0.80 | −0.81 |
| −0.4 | 3 | −0.36 | −0.65 |
| −0.4 | 5 | −0.37 | −0.55 |
| −0.4 | 8 | −0.39 | −0.49 |
| 0.0 | 3 | 0.07 | −0.46 |
| 0.0 | 5 | 0.05 | −0.28 |
| 0.0 | 8 | 0.00 | −0.19 |
| 0.4 | 3 | 0.52 | −0.31 |
| 0.4 | 5 | 0.40 | −0.05 |
| 0.4 | 8 | 0.40 | 0.12 |

Source: `simulation_results_latest.csv`; linear envelope cells only. Separate: Pearson correlation of per-athlete STAR point estimates.

## F.2 Question 1: correlation recovery

Table 3 reports Monte Carlo means of $\hat{\rho}_{\alpha\delta}$ under a linear envelope (the geometry closest to the sprint application). The hierarchical Laplace fit tracks the true association across the sparsity grid. Post-hoc correlations of separate athlete fits are unreliable under sparsity: at $\rho_{\alpha\delta} = 0.4$ they are near zero or wrong-signed for $n_i \leq 5$, while the hierarchical estimand remains close to the truth. Under strongly curved envelopes (quadratic, Gompertz) recovery of $\rho_{\alpha\delta}$ is harder for every method; those cells are in the archive and are not the primary support for the population association claimed in the application.

## F.3 Question 2: geometry and timing

Table 4 summarizes timing recovery at $\rho_{\alpha\delta} = -0.8$. Under an exactly linear envelope, free $\gamma_i$ is not meaningfully recovered: RMSE is large and the correlation with truth is unavailable (matching Proposition 1). Under quadratic and especially Gompertz envelopes, RMSE falls and $\mathrm{Cor}(\hat{\gamma}_i, \gamma_i)$ becomes positive, so finite-sample behaviour tracks the curved case of Proposition 2 beyond the local neighbourhood used in the proof.

## F.4 Question 3: envelope misspecification

Table 5 contrasts a correctly specified Stage 2 template with a geometry-mismatched template (linear truth fit with a curved working envelope, or curved truth fit with a linear working envelope) at $\rho_{\alpha\delta} = -0.8$ and $n_i = 5$. Misspecification attenuates or distorts $\hat{\rho}_{\alpha\delta}$ and inflates timing error when curvature is present in the truth. Matching qualitative geometry—linear versus clearly curved—matters more than modest vertical shifts of a correctly shaped template.

Table 4: Simulation Question 2: timing recovery at $\rho_{\alpha\delta} = -0.8$ ($B = 50$). Entries are Monte Carlo means of RMSE($\hat{\gamma}$) and Cor($\hat{\gamma}, \gamma$).

| Envelope | $n_i$ | RMSE($\hat{\gamma}$) | Cor($\hat{\gamma}, \gamma$) |
|---|---|---|---|
| Linear | 3 | 2.09 | — |
| Linear | 5 | 3.24 | — |
| Linear | 8 | 3.27 | — |
| Quadratic | 3 | 2.87 | 0.09 |
| Quadratic | 5 | 2.23 | 0.22 |
| Quadratic | 8 | 1.77 | 0.14 |
| Gompertz | 3 | 1.64 | 0.43 |
| Gompertz | 5 | 1.38 | 0.50 |
| Gompertz | 8 | 1.39 | 0.50 |

Source: Same Monte Carlo as Table 3. Dashes: correlation not reported when timing is effectively unidentified under the linear truth.

Table 5: Simulation Question 3: correctly specified versus geometry-misspecified Stage 2 at $\rho_{\alpha\delta} = -0.8$, $n_i = 5$ ($B = 50$). Monte Carlo means.

| Truth | $\hat{\rho}$ (correct) | $\hat{\rho}$ (misspec.) | RMSE($\hat{\gamma}$) (correct) | RMSE($\hat{\gamma}$) (misspec.) |
|---|---|---|---|---|
| Linear | −0.79 | 0.03 | 3.24 | — |
| Quadratic | −0.72 | −0.97 | 2.23 | 5.10 |
| Gompertz | −0.14 | −0.97 | 1.38 | 4.88 |

Source: Same Monte Carlo. Misspecification swaps linear and curved working templates as in the simulation script. Dashes: timing RMSE under the reverse misspecification not reported for the linear-truth cell in the archived summary.



\end{document}